\documentclass[aps,pre,preprint,groupedaddress,showkeys,nolinenumbers]{revtex4-1}%twocolumn% showpacs
\usepackage{graphicx}%figure
\usepackage{amsmath}
\usepackage{helvet}%arial font
\usepackage{MnSymbol}%mnsymbol.sty downloaded for \llangle and \rrangle
\usepackage[caption=false]{subfig} %fore subfigure

\usepackage[margin=1in]{geometry}
\usepackage{xifthen}%\ifthenelse{\equal{#2}{}}{empty}{given}
\usepackage{bigints}

\def\sgn{\mathop{\rm sgn}\nolimits} 
\def\lambdah{\hat{\lambda}}
\def\lambdahat{\hat{\lambda}}
\def\zhat{\hat{z}}
\def\sigmahat{\hat{\sigma}}
\def\P{\text{P}}
\def\widthfig{4.4cm} %column width PRL 8.6cm

\newcommand{\me}{\mathrm{e}}%math e
\newcommand\Dx[2]{%
	\ifthenelse{\equal{#2}{}}%
	{\ensuremath{\frac{d#1}{\sqrt{2\pi}}  \, \me^{-{#1^2\over{2}}  }}}%if 2nd argment is empty
	{\ensuremath{{\frac{d#1}{\sqrt{2\pi}} \, \me^{-{#1^2\over{2}}+#2}}}}%if 2nd argment is given
}%usage: $\Dx{t}{bt}$
\newcommand{\icmplx}{\mathrm{i}}
\newcommand\ddfrac[2]{\frac{\displaystyle #1}{\displaystyle #2}}

\def\Mhat{\hat{M}}

\usepackage[nameinlink]{cleveref} %for \cref{eq:1,eq:2} \crefrange{first}{last}
\crefname{equation}{eq.}{eq.}

\crefname{ineq}{inequality}{inequalities} %\label[ineq]{i:positivity} %\Cref{i:positivity}
\creflabelformat{ineq}{#2{\upshape(#1)}#3}

\crefname{cond}{condition}{conditions} %\label[cond]{cond:mainconstraint} %\Cref{cond:mainconstraint}
\creflabelformat{cond}{#2{\upshape(#1)}#3}

\crefname{noname}{}{} %\label[noname]{cond:mainconstraint} %\Cref{cond:mainconstraint}
\creflabelformat{noname}{#2{\upshape(#1)}#3}

\begin{document}

%\preprint{}

\title{Exponential Capacity in an Autoencoder Neural Network with a Hidden Layer}


\author{Alireza Alemi}
\email[Corresponding author: ]{alireza.alemi@\{ens.fr,\hspace{1ex}gmail.com\}}
\author{Alia Abbara}
%\thanks{}
\affiliation{Group for Neural Theory, \'Ecole Normale Sup\'erieure, 29 Rue d'Ulm -- 75005, Paris, France}


\date{\today}

\begin{abstract}
 %Neural network models learn different tasks by 
 
 A fundamental aspect of limitations in learning any computation in neural architectures is characterizing their optimal capacities. 
 %arising from adapting their parameters. 
 An important, widely-used neural architecture is known as autoencoders where the network reconstructs the input at the output layer via a representation at a hidden layer.
  Even though capacities of several neural architectures have been addressed using statistical physics methods, the capacity of autoencoder neural networks is not well-explored.
 Here, we analytically show that an autoencoder network of binary neurons with a hidden layer can achieve a capacity that grows exponentially with network size. 
 The network has fixed random weights encoding a set of dense input patterns into a dense, expanded (or \emph{overcomplete}) hidden layer representation. A set of learnable weights decodes the input patters at the output layer. We perform a mean-field approximation of the model to reduce the model to a perceptron problem with an input-output dependency. Carrying out Gardner's \emph{replica}  calculation, we show that as the expansion ratio, defined as the number of hidden units over the number of input units, increases, the autoencoding capacity grows exponentially even when the sparseness or the coding level of the hidden layer representation is changed. The replica-symmetric solution is locally stable and is in good agreement with simulation results obtained using a local learning rule. In addition, the degree of symmetry between the encoding and decoding weights monotonically increases with the expansion ratio.

\end{abstract}

%89.75.Fb complex systems: Structures and organization in complex systems

\maketitle

%\section{}
%\subsection{}
%\subsubsection{}

Characterizing the power and limitations of neural network architectures for performing different computations is an important step toward understanding any neural systems. Network architectures with hidden layers provide very powerful computational power both in artificial \cite{lecun15,goodfellow16} and biological neural systems \cite{cadieu14}.
An important class of neural networks with hidden layers, known as autoencoders, reconstructs back the input patterns at the output layer via a code at a hidden layer that represents the input. By applying constraints such as sparseness on the hidden layer, they provide useful representation of the input for a variety of tasks. The autoencoding capacity is determined by the number of hidden units; however, a general theory for this capacity has not been put forward.
 Furthermore, a theory on the capacity of this architecture can open a door for better understanding deep neural networks with several hidden layers which have dramatically improved the performance of machine learning systems in a variety of tasks such as visual object recognition and speech recognition tasks \cite{lecun15}.

 Traditionally, statistical physics methods have been extensively used to characterize capacities of neural networks for classification and generalization problems, and similar related problems \cite{gardner88,engel01,advani13}. The maximal storage capacity per synapse in a simple model neuron, known as the simple perceptron, has been calculated by Elizabeth Gardner using her \emph{replica} theory showing the capacity is $\alpha_c=2$ \cite{gardner88}.
 The method has been widely applied to a variety of cases in perceptrons such as binary weights \cite{krauth89}, generalization problem \cite{engel01}, and spatially-correlated patterns \cite{monasson92,monasson93}.
 Furthermore, the study of optimal storage properties of the perceptron and recurrent networks based on Gardner's method has provided parsimonious theories for statistics of synaptic weights in neural circuits \cite{brunel04,chapeton12,chapeton15,brunel16}. 
In comparison with the simple perceptron, the Gardner analysis is more complicated for network architectures with hidden layers because of complexities arising from additional internal degrees of freedom in the hidden layers. Consequently, researchers turned to studying tailored architectures such as the committee machine or the parity machine, which are amenable to Gardner analysis. For example, for the fully-connected committee machine, with $N\rightarrow\infty$ input neurons, $K$($<<N$) hidden neurons and $p$ input patterns, the critical capacity $\alpha_c=p_c/N$ scales as $K\sqrt{\log K}$ \cite{urbanczik97,kwon97}. 
In spite of such progress, the maximal capacity of networks with hidden layers in the general case has not been tackled yet. In particular, the Gardner replica method has not been applied to autoencoders which solve a different problem than multilayer perceptrons (MLP) in spite of having hidden layer representations.

In this Letter, we attempt at exploring the capacity of a simple autoencoder that is analytically tractable: the network has random encoding weights and an expansive (or \emph{overcomplete}) architecture which provides an unexpected capacity.  
Thanks to the random projection and a mean-field approximation (MFA), we reduce the problem to computing the capacity of a perceptron using Gardner's replica method. The inputs and the output of the perceptron have very small correlations resulting in an unexpected result: we find that the capacity grows exponentially with the ratio of the number of hidden units to the number of input units.

 \begin{figure}[b]
	%TODO: change synapse to weight or connection and plastic to learnable/modifiable
	\centering
	\includegraphics[width=7cm]{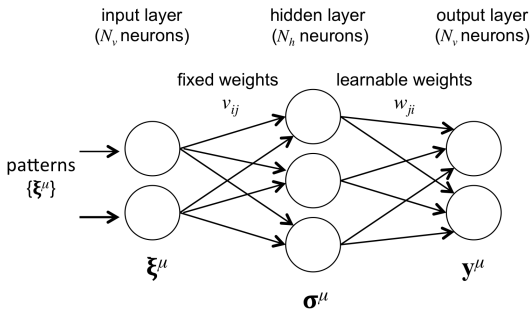}%scale=0.12
	%	\includegraphics{model.png}%scale=0.12
	%\internallinenumbers
	\caption{
		\label{model} Autoencoder neural network structure. The network has three layers: the input with $N_v$ neurons, the hidden with $N_h$ neurons, and the output layers with the same number of neurons as the input layer. The output, after learning, reconstructs back the input patterns $\mathbf{y}^\mu=\boldsymbol\xi^\mu$ via the hidden layer representation $\boldsymbol{\sigma}^\mu$. The input patterns are projected to the hidden layer by a set of fixed random encoding connections 
		$v_{ij}$.
		The decoding connections are learned to reproduce the input patterns at the output layer.
	}
\end{figure}

%% The Model
Our autoencoder model is made of three layers: an input layer ($N_v$ neurons), a hidden layer ($N_h$ neurons), and an output layer ($N_v$ neurons), as seen in Fig. \ref{model}. We consider the McCulloch-Pitts neuron model with the sign transfer function --- defined as $\sgn(x)=+1$ if $x>0$ and $\sgn(x)=-1$ otherwise --- and the synaptic input is summed linearly. The input and output layers have the same number of neurons. The input patterns are encoded into the hidden layer representation. The patterns need to be decoded from this representation at the output layer. The network is fully-connected between layers with no lateral connectivity within layers. The binary ($\pm1$) input patterns $\{\boldsymbol\xi^\mu\}$, 
%$\{\xi^\mu=\xi^1_1...\xi^n_j...\xi^p_N_v\}$, 
where $\mu=1,2,..., p$, are presented to the input layer. For a given pattern $\boldsymbol{\xi}^\mu$, each input layer neuron $\xi_j^\mu$, where $j=1,2,...,N_v$, projects with fixed random connections $v_{ij}$ to the hidden layer neurons denoted by $\sigma_i^\mu$ where $i=1, 2, ..., N_h$. The encoding weights $v_{ij}$ are sampled from a normal distribution ${\cal{N}}(0,1)$ and remain unchanged during learning process. The decoding weights $w_{ji}$ are initialized randomly but learned during the training phase. The goal of learning is that for each pattern $\mu$ the output layer neurons $y_j^\mu$ reproduce $\xi_j^\mu$ after learning is done i.e. $y_j^\mu=\xi_j^\mu$. We used an online learning rule, the perceptron learning rule (PLR), in the simulations \cite{supplim}. The entries of the pattern matrix $\xi_j^\mu$ are independent and identically distributed random variables with probability distribution $\text{P}(\xi_j^\mu=\pm1)=0.5$ yielding dense regime patterns ($0.5$ coding level).

We define the expansion ratio $\Lambda=N_h/N_v$, and we are mainly interested in expansive autoencoders with ${\Lambda\geq 1}$, where the exponential capacity occurs. The maximal capacity ratio (or simply the maximal capacity) is defined as the maximal number of patterns that can be decoded at the output layer divided by the number of hidden layer units, i.e. $\alpha_c=p_{max}/N_h$. We are interested in calculating the maximal capacity in the thermodynamic limit $N_v,N_h\rightarrow\infty$ but with finite $\Lambda$.

Once a pattern $\boldsymbol\xi^\mu$ is presented at the input layer, the network will update the corresponding hidden layer $\boldsymbol\sigma^\mu$ and the output layer $\mathbf{y}^\mu$ using the following dynamics for each $i$ and $j$: $\sigma_{i}^\mu = \sgn \big(\sum_{j=1}^{N_v} v_{ij}\xi_{j}^\mu\big)$  and $y_{j}^\mu = \sgn \big(\sum_{i=1}^{N_h} w_{ji}\sigma_{i}^\mu\big)$. Since the patterns are unbiased the neuronal threshold is considered to be zero in the dynamics equations. The hidden units operate at $0.5$ coding level due to the encoding weights being random with zero mean.  Without loss of generality (w.l.o.g), we consider the spherical constraint $\sum_{i=1}^{N_h}w_{ji}^2=N_h$ for each $j$ as it has no effect on learning the decoding weights. It is desired in an autoencoder model that $y_j^\mu=\xi_j^\mu$ for each $j$ and $\mu$. This requires that, in order to perfectly reconstruct the patterns, the following conditions must hold for each $\mu$ and $j$: 
\begin{equation}
\xi_j^{\mu} = \sgn \big( \sum_i w_{ji} \sgn \big(\sum_l v_{il} \xi_l^{\mu}\big)\big).
\label{eq:fixedpoint}
\end{equation}
In order to make the Eq.~(\ref{eq:fixedpoint}) amenable to Gardner's replica calculations, we propose the following mean-field approximation (MFA) where we separate the contribution of $\xi_j^\mu$, w.l.o.g. for an arbitrary chosen $j$, in the local field of the hidden units and treat the rest of the summation as quenched Gaussian noise $z_i^\mu$. The hidden unit dynamic then becomes 
\begin{align}
	\sigma_{i}^\mu = \text{sgn} \Big( \sum_{l\neq j;l=1}^{N_v} v_{il} \xi_l^{\mu}+v_{ij}\xi_j^\mu \Big)
	% \sgn \big(\sum_{j=1}^{N_v} v_{ij}\xi_{j}^\mu\big)
	\approx\text{sgn}\Big(z_i^\mu+v_{ij}\xi_j^\mu\Big)
\end{align}
%\begin{align}
%	\xi_j^\mu	&= \text{sgn} \bigg( \sum_{i=1}^{N_h} w_{ji}\, \text{sgn} \Big(\sum_{l\neq j;l=1}^{N_v} v_{il} \xi_l^{\mu}+v_{ij}\xi_j^\mu\Big)\bigg)\nonumber\\
%	&= \text{sgn} \bigg( \sum_{i=1}^{N_h} w_{ji}\, \text{sgn} \Big(z_i^\mu+v_{ij}\xi_j^\mu\Big)\bigg),
%	\label{eq:mfa}
%\end{align}
where the random variable $z_i^\mu\sim {\cal{N}}(0,N_v)$ in the limit of large $N_v\rightarrow\infty$. This approximation can be viewed as following: in finding the decoding weights for an arbitrary element $\xi_j^\mu$,  all higher-order correlations $\llangle\sigma_{i_1}^\mu...\sigma_{i_L}^\mu\xi_j^\mu\rrangle$
can be expressed as a function of the pairwise correlations $\llangle\sigma_i^\mu\xi_j^\mu\rrangle$ where $\llangle.\rrangle$ denotes average over the ensemble of all pattern matrices \cite{supplim}. Therefore, we have discarded some of the complexities of the correlation structure in the full model by introducing the MFA.
%This means that the MFA has less complexity with  in terms correlations structure. 
%This makes all pairs of hidden units conditionally independent given $\xi_j^\mu$ i.e. $\P(\sigma_i^\mu \,\vert \, \xi_j^\mu) \perp \P(\sigma_k^\mu \, \vert \, \xi_j^\mu)$. 
It must be noted that due to replacing the higher order correlation with the Gaussian noise in the decoding process, the hidden units $\boldsymbol{\sigma}^\mu$ carry less decodable information about $\xi_j^\mu$ 
in the MFA than in the full network causing a decrease of capacity in the MFA with respect to the full-network capacity. But as we will see, the exponential capacity can still be captured in the MFA model. We should note that, unlike the classical mean-field theories where increasing the system size makes the calculation more exact, here increasing the system size cannot recover the loss in the capacity.

The MFA model can now be reformulated as a perceptron problem with $\boldsymbol{\sigma}^\mu$ as its input and $\xi_j^\mu$ as its output label. 
%In order to do so, we note that whenever the conditions 
%\begin{equation}
%\forall i:\, \lvert z_i^\mu\rvert<\lvert v_{ij}\rvert \text{ \hspace{2ex} and \hspace{2ex} } \sum_{i=1}^{N_h} w_{ji}v_{ij}>0 
%\label{eq:conditionW}
%\end{equation}
%are met, the Eq.~(\ref{eq:mfa}) hold. 
This allows us to compute the conditional probabilities  \cite{supplim} 
\begin{equation}
\P\big(\sigma_i^\mu \big\vert \xi_j^\mu\big)\simeq\frac{1}{2}+\sigma_i^\mu\xi_j^\mu\frac{v_{ij}}{\sqrt{2\pi N_v}}. \label{eq:prob4}
\end{equation}

We can implement sparseness by changing the fraction of active neurons, $f$, in the hidden layer, ensuring that $\text{P} (\sigma_i^\mu=+1) = f$ and ${\text{P}(\sigma_i^\mu=-1)=1-f}$. This is done by adding a threshold $\theta$ in the transfer function of the hidden units  which becomes
				%\begin{equation}
				${\sigma_{i\text{, sparse}}^\mu = \sgn \big(\sum_{j=1}^{N_v} v_{ij}\xi_{j}^\mu - \theta\big)}$,
				%\end{equation}
where ${\theta=\sqrt{N_v}H^{-1}(f)}$, $H(x)\equiv \int_x^{\infty} \ddfrac{dt}{\sqrt{2\pi}}\me^\frac{-t^2}{2}$, and $H^{-1}(.)$ is the inverse function of $H(.)$. The MFA can be applied to the sparse case \cite{supplim},  yielding conditional probabilities:
\begin{equation}
%\P\big(\sigma_i^\mu \big\vert \xi_j^\mu\big)\simeq\frac{1}{2}+\sigma_i^\mu\xi_j^\mu\frac{v_{ij}}{\sqrt{2\pi N_v}}-\sigma_i^\mu \frac{H^{-1}(f)}{\sqrt{2\pi}} 
\P\big(\sigma_{i}^\mu \big\vert \xi_j^\mu\big)\simeq\frac{1}{2}+\sigma_i^\mu\xi_j^\mu\frac{v_{ij}}{\sqrt{2\pi N_v}}\exp\left( - \frac{[H^{-1}(f)]^2}{2}\right).
\label{eq:prob_sparse}
\end{equation}

%Replica theory
	Following standard Gardner's replica calculation \cite{gardner88}, we need to calculate the typical volume of solutions of our perceptron in the weight space in the thermodynamic limit where the dimensionality of layers $N_v,N_h\rightarrow\infty$ and the number of patterns $p\rightarrow\infty$ with finite $\alpha=p/N_h$ and $\Lambda=N_h/N_v$. At maximal capacity, the typical volume shrinks to a unique solution.  In order for a pattern indexed $\mu$ to be a solution of the perceptron and be robust with a margin, we enforce the following requirement
	\begin{equation}
	\xi_j^\mu\Big(\frac{1}{\sqrt{N_h}}
	\sum_{i=1}^{N_h}w_{ji}\sigma_i^\mu \Big) > \kappa 
	%\label{eq:mainconstraint}
	\label{cond:mainconstraint}
	\end{equation}
	where $\kappa$ is a robustness parameter providing a margin for the solution --- the larger the $\kappa$, the larger the margin.
	The Gardner volume, for a given realization of $\xi_j^\mu$, $\sigma^\mu$,$v_{ij}$ and for a fixed $j$, is
	\begin{equation}
	\Omega = {\int_{\| \mathbf{w} \|^2=N_h} d^{N_h}\mathbf{w}\,
		\prod_{\mu=1}^{p} 
		\Theta\big( \xi_j^\mu \frac{1}{\sqrt{N_h}}\sum_{i=1}^{N_h}w_{ji}\sigma_i^\mu - \kappa \big)},
	\end{equation}
   where $\Theta(.)$ is the Heaviside step function, $\kappa$ is a robustness parameter. Assuming the volume is self-averaging (as in \cite{gardner88}),  we only need to calculate the quenched average 
$\llangle\log(\Omega)\rrangle_{\xi_j^\mu,\boldsymbol\sigma^\mu,\mathbf{v}_{j}}$ where  ${\mathbf{v}_j\equiv v_{.j}}$ with the spherical constraint has %$\|\mathbf{v}_j\|^2=N_h$. 
%$\|\mathbf{v}_j\|^2=\sum_i v_{ij}^2=N_h$. 
the distribution ${\P(\mathbf{v}_{j})=	(2\pi\me)^{-N_h/2}\delta(\|\mathbf{v}_{j}\|^2-N_h)}$. Note that the difference between our calculation and standard Gardner calculation is that we have a dependency between the input $\boldsymbol{\sigma}^\mu$ and the output $\xi_j^\mu$ of the reduced perceptron, which is given by the conditional probability distribution Eq.~(\ref{eq:prob4}) in the general case, and Eq.~(\ref{eq:prob_sparse}) in the sparse case. Using the replica method, the problem is transformed into calculating the quenched average of $\llangle\Omega^{n}\rrangle_{\xi_j^\mu,\boldsymbol{\sigma}^\mu,\mathbf{v}_{j}}$ related to $n$ replicas of the system and taking the limit of $n\rightarrow 0$. We use the replica-symmetric (RS)
 ansatz which  is known to give the correct result for the capacity of simple perceptrons with  continuous weights where the space of solutions is connected therefore the replica method is known to yield correct results. We show that RS solution is locally stable \cite{supplim}.
%\floatsetup[figure]{style=plain,subcapbesideposition=top}
  \begin{figure}[b]

	\centering
	\begin{minipage}{\linewidth}
		\subfloat[]{\includegraphics[width=\widthfig]{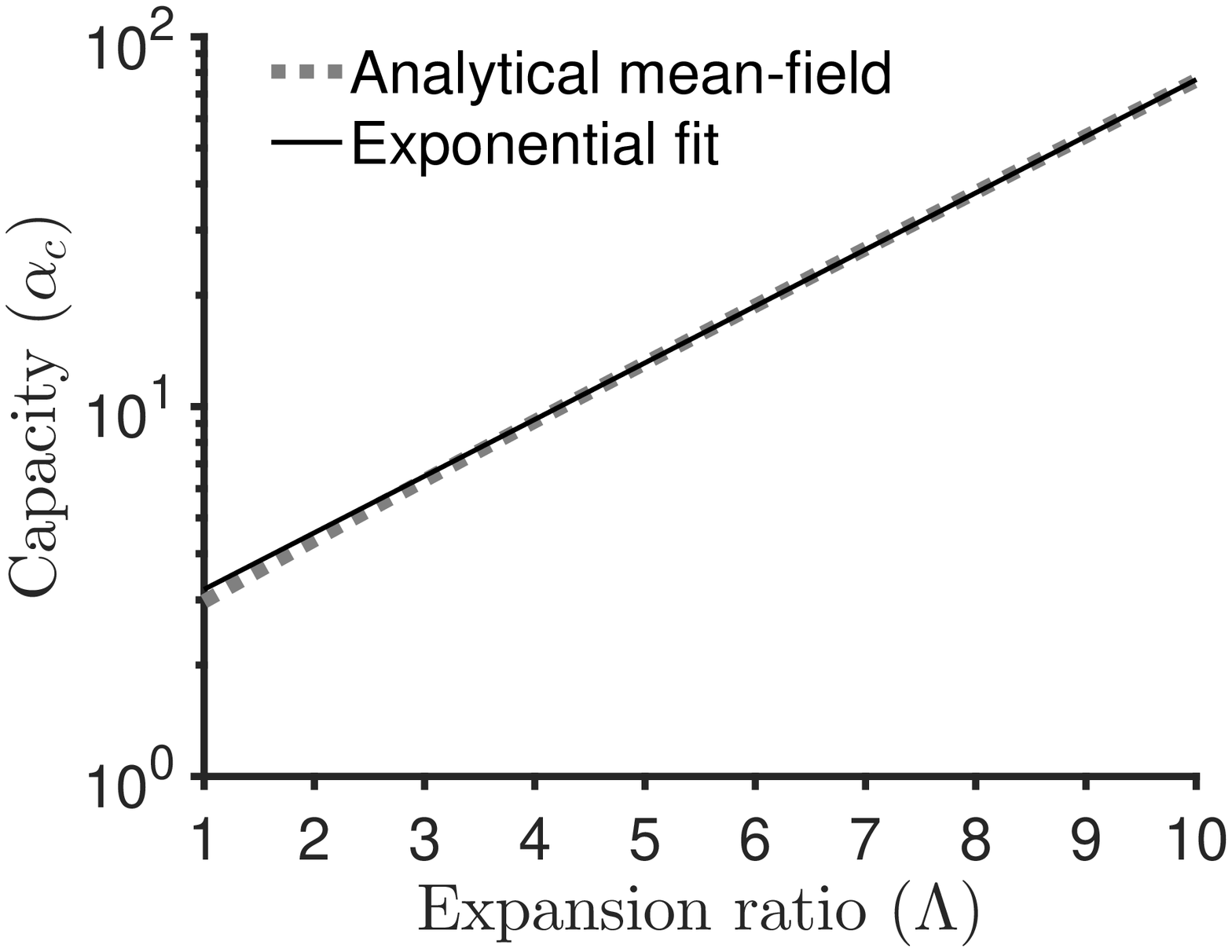}}
		 \subfloat[]{\includegraphics[width=\widthfig]{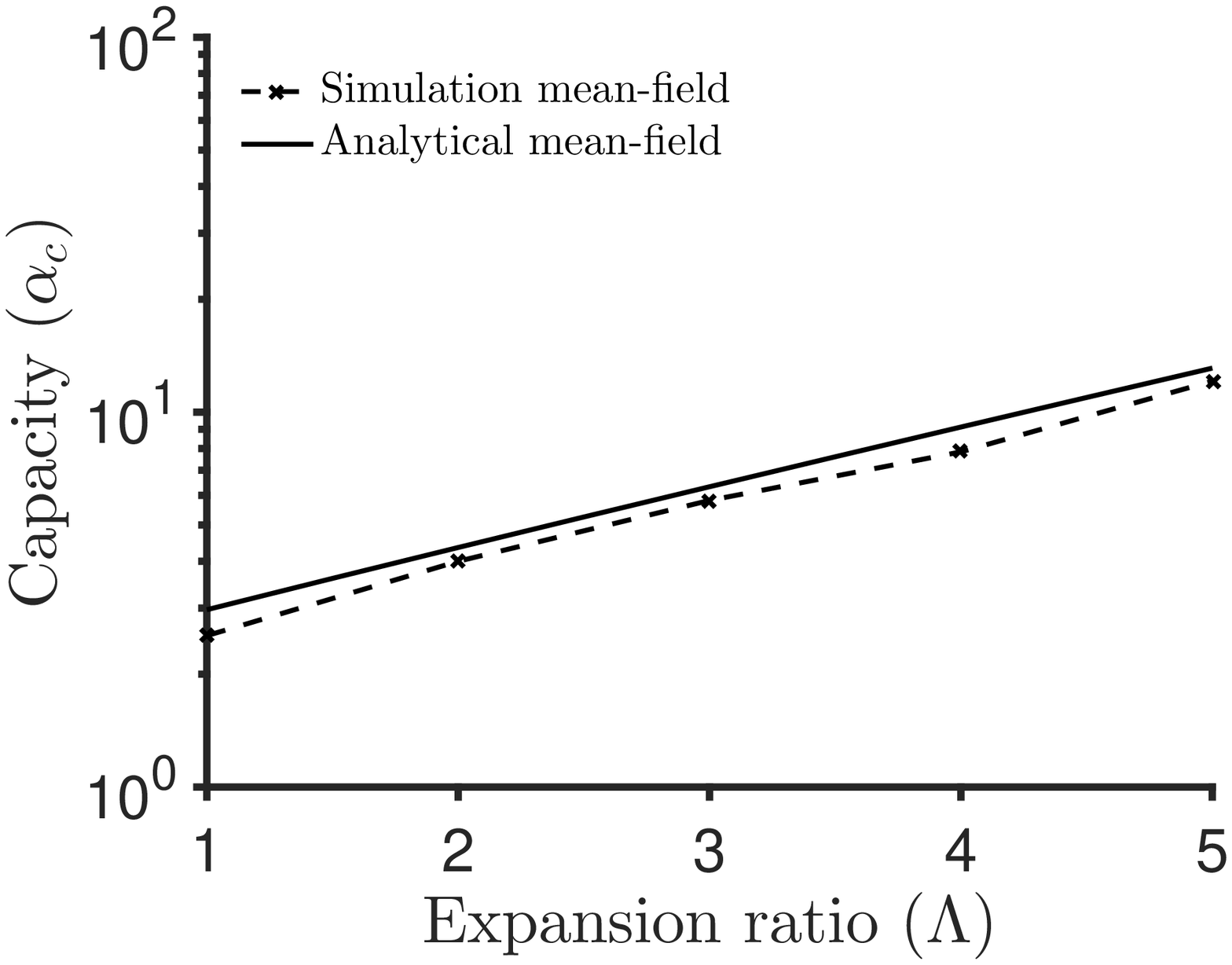}}
	\end{minipage}
	
	\begin{minipage}{\linewidth}
		\subfloat[]{\includegraphics[width=\widthfig]{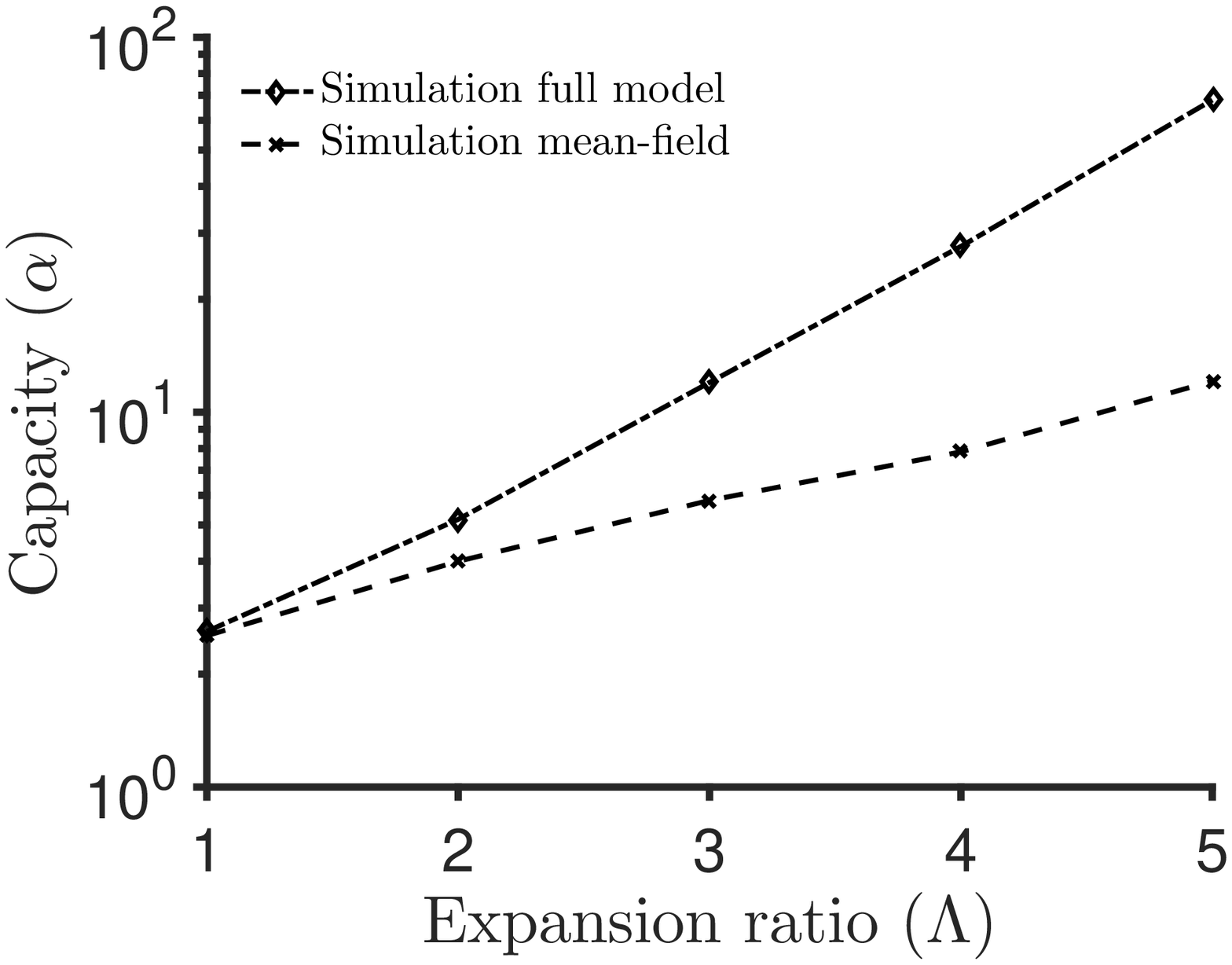}}
		\subfloat[]{\includegraphics[width=\widthfig]{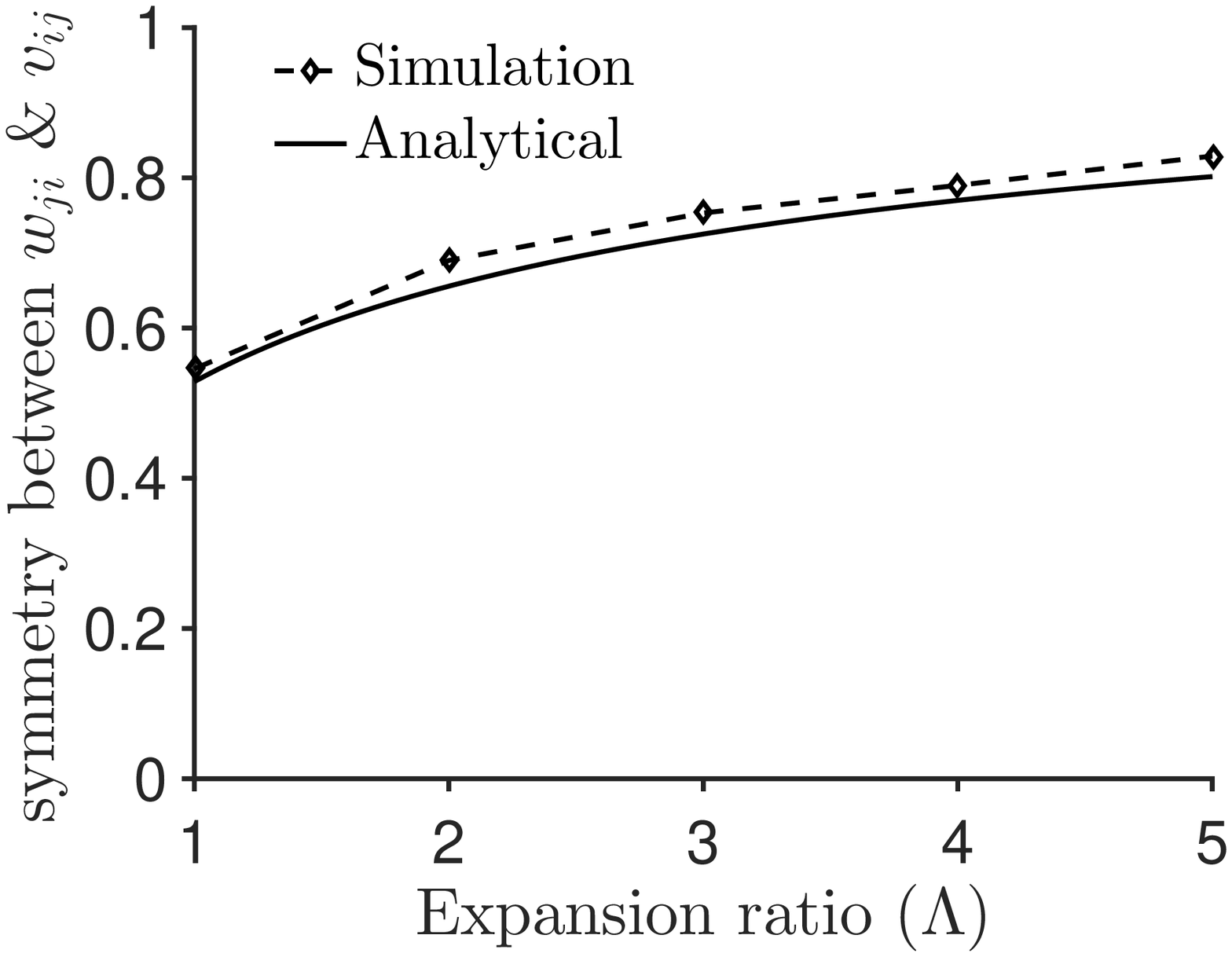}}
	\end{minipage}
  	%\internallinenumbers
	\caption{\label{fig:analytCapacity} Comparison between analytical and simulation results.~~(a) An exponential fit to the analytical mean-field result at $\kappa=0$ for $\Lambda\in[1,10]$ shows that the growth can be approximated with a small error with the exponential expression $15.66\,\me^{0.9239\Lambda}$ with the fitting error $\text{RMSE}=0.1413$. (b)~The capacity is plotted as a function of the expansion ratio on a semi-log scale for simulation and analytical mean-field model at $\kappa=0$.  (c) Due to the simplification of higher order correlations after introducing the mean-field model, the capacity of the mean-field is lower than the full-model. This is illustrated by running simulations for $N_v=100$ number of neurons. (d) The plot compares the simulation and analytical results for the symmetry between the encoding and decoding weights. As the the expansion ratio increases, this symmetry monotonically increases.}
\end{figure}

%% Result
We find \cite{supplim} that the following two integral equations determine the critical capacity in the general case:
  \begin{eqnarray}
\alpha_c^\text{MFA} \int_{M\sqrt{2 \Lambda/\pi} - \kappa}^{\infty} Dt \, \big( \kappa + t - \sqrt{2 \Lambda/\pi} \big)^{2} = 1- M^{2}\label{eq:result1} \hspace{2ex}\\
\alpha_c^\text{MFA} \int_{M\sqrt{2 \Lambda/\pi}-\kappa}^{\infty} Dt \, \big( \kappa + t - \sqrt{2 \Lambda/\pi} \big) \sqrt{2 \Lambda/\pi} = M \hspace{2ex}
\label{eq:result2}
\end{eqnarray}
 where $\alpha_c^\text{MFA}$ is the critical capacity of the MFA, ${M\equiv M_j=\sum_i\ddfrac{v_{ij} w_{ji}}{N_h}}$ is the degree of symmetry between the encoding and decoding connections, and ${Dt \equiv \ddfrac{dt}{\sqrt{2\pi}}\me^\frac{-t^2}{2}}$. These equations can be solved numerically, showing, at $\kappa=0$, the critical capacity grows as a function of $\Lambda$ that can be approximated by an exponential $a \exp(b\Lambda)$
  with $a = 15.66$, with confidence interval $(15.64, 15.69)$, and $b= 0.9239$ with confidence interval
  	 $(0.9226, 0.9252)$ and the fitting error ${\text{RMSE}=0.1413}$ (see Fig.~\ref{fig:analytCapacity}(a)). In the limit $\Lambda\rightarrow\infty$, this capacity scales as $\alpha_c^\text{MFA} \sim \sqrt{\Lambda}\me^{\Lambda/\pi}$.  
  	 %In the limit $\Lambda\rightarrow\infty$, the scaling is  ${\alpha_c \sim \sqrt{\Lambda}\exp{(\frac{\Lambda}{\pi})}}$. In the limit $\Lambda\rightarrow0$, the critical capacity $\alpha_c\rightarrow2$. 
 \begin{figure}[]
 	\centering
 	\subfloat[]{\includegraphics[width=4.0cm]{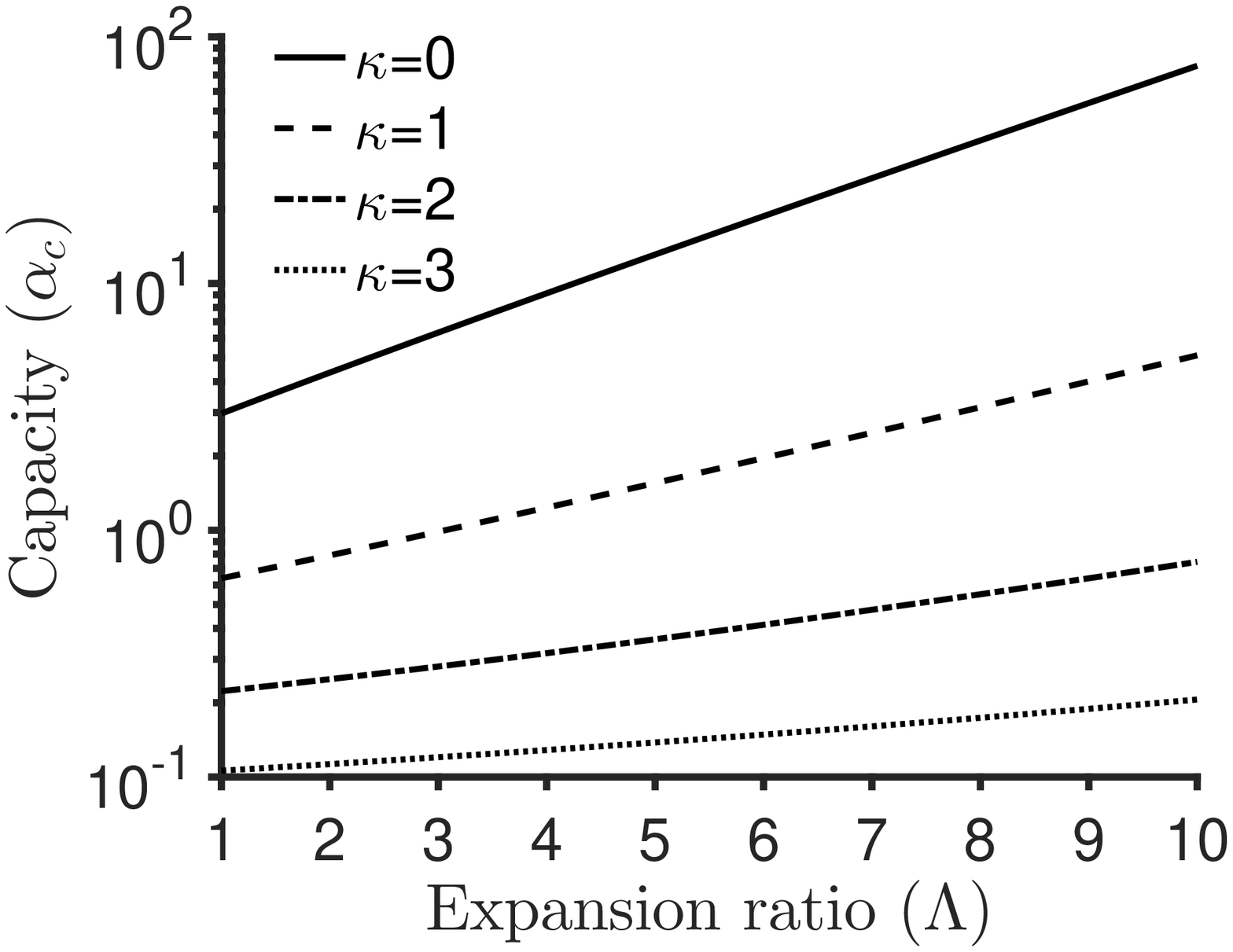}} %column width PRL 8.6cm
 	\subfloat[]{\includegraphics[width=4.0cm]{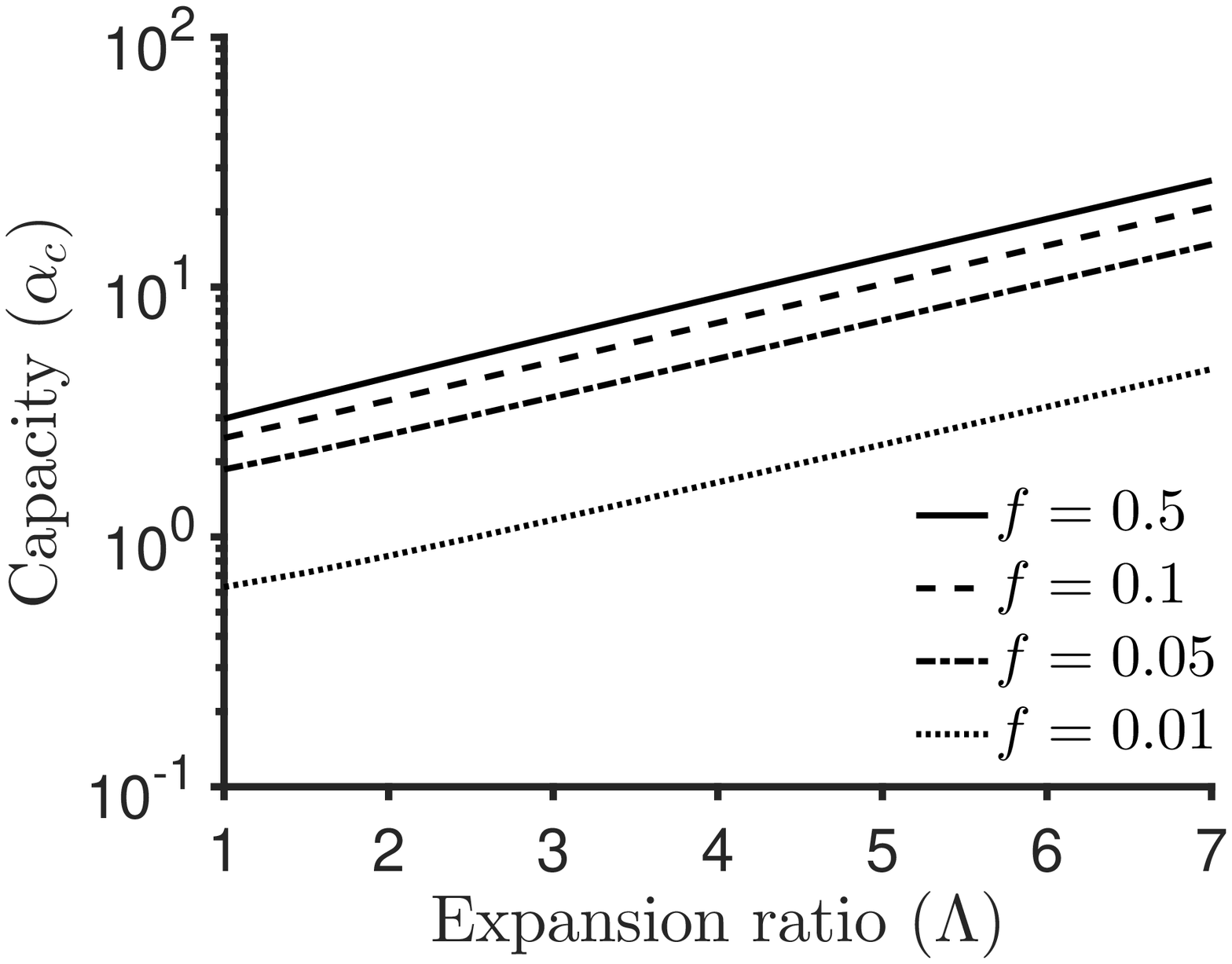}}
 	%\internallinenumbers
 	\caption{\label{fig:kappa} The effect of robustness to noise and sparseness in the hidden units on the critical capacity of the MFA.~~(a)~The capacity as a function of expansion ratio is plotted for a range of values of robustness $\kappa$ in a semi-log scale. The growth of capacity is exponential for all the shown values of $\kappa$. (b)~Analytical results showing the capacity grows exponentially with $\Lambda$ for different levels of sparseness in the hidden units.}
 \end{figure}
 The result of the simulations of the MFA with ${N_v=100}$ \cite{supplim} is in good agreement with the analytical result suggesting an exponential growth of capacity with the expansion ratio $\Lambda$ on a semi-log scale plot as shown in Fig.~\ref{fig:analytCapacity}(b).
  
  This surprising exponential capacity of the MFA model is due to $N_h$ number ($\rightarrow\infty$) of vanishing small, pairwise correlations between the input and the output units, which are denoted by $m_i = \llangle \sigma_i^\mu \xi_j^\mu \rrangle = \frac{2v_{ij}\sqrt{\Lambda}}{\sqrt{2\pi N_h}}$ in our reduced perceptron. Infinite number of very small pairwise correlations provides strong evidence for the perceptron to perform the input-output association task.% with the given probability distributions. 
  
  The capacity in the MFA does not grow as fast as the capacity in the full model with the same network size $N_v=100$ as compared by simulations in Fig~\ref{fig:analytCapacity}(c). The reason that the full model has a higher capacity (higher exponent in the exponential growth) than that of the MFA is due to the full model having more complex higher-order correlations than in the MFA.

  Our theory also shows that there is a relation between the  weights $v_{ij}$ encoding $\xi_j^\mu$ into the hidden layer representation and the weights $w_{ji}$ which are trying to decode $\xi_j^\mu$ from that representation. This is quantified as the degree of symmetry between the $v_{ij}$'s and the  $w_{ji}$'s for a network that operates at the critical capacity  and is shown in Fig.~\ref{fig:analytCapacity}(d), for the simulation and analytical results. We observe that, at critical capacity, the network becomes more symmetric as the expansion ratio increases and in the limit of $\Lambda\rightarrow\infty$ the network becomes fully-symmetric.

% \begin{figure}
% 	\centering
% 	\includegraphics[width=4.3cm]{alpha_lambda_simulOnly_BW}
% \end{figure}
 
 %Capacity as a function of total number of neurons  in \ref{fig:sim}B to show the rapid increase with network resources. For comparison, the Gardner bound for a fully-connected recurrent neural network with the same number of neurons is plotted. The increase in capacity is beyond the Gardner bound which is due to the presence of correlated hidden units, whereas the Gardner bound in a recurrent neuron network refers to the case that neurons are not correlated. Note that the number of learnable weights for the orange curve, i.e. $(N_v+N_h)(N_v+N_h-1)/2$, is much higher than the number of learnable weights in the autoencoder, i.e. $N_v N_h$.
 %\renewcommand*{\familydefault}{\sfdefault}

  Until now, we have focused on the results at zero robustness but the growth of capacity is still exponential when we consider a margin for the solution so that the solutions are robust to bit flips of the hidden units or white noise added to the decoding weights. We make our solutions robust to these noises by increasing $\kappa$. Fig.~\ref{fig:kappa}(a) compares the analytical MFA capacity for $\kappa=0$ and non-zero robustness $\kappa$. Though the slope of the line in a semi-log scale gets smaller as we increase $\kappa$, the growth is still very close to exponential for non-zero $\kappa$. The robustness to a bit flip in the input layer might be harder to obtain while achieving an exponential capacity. Preliminary arguments \cite{supplim} for the MFA show that by setting $\kappa$ large enough to be robust to a bit flip in the input, the capacity decreases with $\Lambda$.
  
   In the sparse hidden units regime,
   %where ${\text{P} (\sigma_i^\mu=+1) = f}$ and $\text{P}(\sigma_i^\mu=-1)=1-f$, 
   Eqs.~\eqref{eq:result1} and \eqref{eq:result2} remain the same except for $M$ which is replaced by ${\tilde{M}=\exp(-\frac{H^{-1}(f)^2}{2})M}$. For fixed values of sparseness $f$, the capacity still grows exponentially as $\Lambda$ grows. The capacity at fixed $\Lambda$ decreases as sparseness increases, as shown in Fig.~\ref{fig:kappa}(b).

  %%Discussion
  In summary, using Gardner's replica method and a mean-field approximation, we derived the capacity of an expansive autoencoder neural network in the MFA. This capacity
   appears 
  to be, according to simulations, a lower bound for the capacity of the full autoencoder. The small correlation between the hidden units and the output shows that the coding that happens in the hidden units is essential in achieving the exponential capacity. 
  %The information is not just stored in the decoding weights but it is first encoded in a representation (hidden units) that yields very small correlation with the readout, helping the decoding weights to be learned.
   This gives a deeper understanding of the role of expansive hidden layers in neural network architectures. The spatial correlation between $\xi_j^\mu$ and $\xi_k^\mu$ across the $\mu$'s is shown to have no effect on the storage capacity of a perceptron \cite{monasson92}. However, this correlation does increase the storage capacity of a recurrent neural network working as an autoassociative memory though not in an exponential fashion \cite{monasson93}.  By structuring patterns, one may store exponential patterns in recurrent neural networks \cite{hillar14,fiete14}.  The autoencoder considered in our study is non-recurrent, but the calculations hold also for the recurrent autoencoder version. %with zero robustness in the input.

  It would be interesting to see how optimizing the encoding weights, diluted connectivity, or adding more hidden layers can affect the trade-off between the capacity and the robustness to noise in the input layer. We used the perceptron learning rule (PLR) for learning in the simulation, but an approximate of PLR, known as the 3TLR \cite{alemi15} can yield similar results without relying on an explicit `error signal' to learn the decoding weights.

   There are theories that study various aspects of networks with hidden layers in special conditions and mainly in low capacity regimes using random connectivity or given specific learning rules \cite{rigotti10,barak13,rigotti14,poole16,kadmon16}. Our case is different, as it studies the capacity of an autoencoder architecture in the optimal scenario which does not depend on the choice of the learning rule. Extension of our framework to deep autoencoders and feedforward networks used for classification is also of great interest and needs to be investigated in future.
  
  A.Al. would like to thank Nicolas Brunel, Sahar Pirmoradian, Carlo Baldassi, Sophie Deneve, Alexis Dubreuil, Peter Latham, Gianluigi Mongillo, Stefano Fusi for useful discussion and/or comments.
  We also acknowledge funding from Agence Nationale de la Recherche (ANR) grant ANR-10-LABX-0087 IEC and ANR-10-IDEX-0001-02 PSL,
  European Research Council (ERC) grant `Predispike', and James S. McDonnell Foundation.

\pagebreak
\widetext

\begin{center}
	\textbf{\large Supplementary Materials (Exponential Capacity in an Autoencoder Neural Network with a Hidden Layer)}
		\text{Alireza Alemi, Alia Abbara}
\end{center}
%%%%%%%%%% Merge with supplemental materials %%%%%%%%%%
%%%%%%%%%% Prefix a "S" to all equations, figures, tables and reset the counter %%%%%%%%%%
\setcounter{equation}{0}
\setcounter{figure}{0}
\setcounter{table}{0}
\setcounter{page}{1}
\makeatletter
\renewcommand{\theequation}{S\arabic{equation}}
\renewcommand{\thefigure}{S\arabic{figure}}
\renewcommand{\bibnumfmt}[1]{[S#1]}
\renewcommand{\citenumfont}[1]{S#1}
%%%%%%%%%% Prefix a "S" to all equations, figures, tables and reset the counter %%%%%%%%%%

%
%
\section{Network Simulation}

The goal of the model is to store a set of $p$ uncorrelated, binary ($\pm1$) patterns $\{\boldsymbol{\xi}^\mu\}$ (where ${\mu\in\{1,2,...,p\}}$) as fixed-points of $\xi_j^{\mu} = \sgn \big( \sum_{i=1}^{N_h} w_{ji} \sgn \big(\sum_{l=1}^{N_v} v_{il} \xi_l^{\mu}\big)\big)$ for each $\mu$ and $j$. The binary variables $\xi_j^\mu$ are independent from each other and are in the dense regime, i.e. with probability ${P(\xi_i^\mu=\pm1)=0.5}$ . The fixed encoding weights $v_{ij}$ are sampled from a Gaussian distribution with mean zero and standard deviation one, ensuring the hidden units work in the dense regime as well. On the other hand, the plastic weights $w_{ji}$ are modified during the learning process. We simulated a synchronous update of the dynamics with discrete time.

The learning rule for updating the encoding weights $w_{ji}$ (the weights are continuous with real value)  is the online version of the perceptron learning rule (PLR). The simulation of the 3TLR yielded very similar results. The learning procedure is as follows. Once a pattern $\mu$ is presented to the input layer, the hidden layer, and the output layer are updated according to the network dynamics. Then the weights are updated using 
%If $\xi_j^\mu\Big(\frac{1}{\sqrt{N_h}} \sum_{i=1}^{N_h}w_{ji}\sigma_i^\mu \Big) < \kappa $, then the  $w_{ji}$'s are updated according to

\begin{equation}
	%\Delta w_{ji} = \eta \xi_j^\mu  \sigma_i^\mu,
	\Delta w_{ji} = \eta (\xi_j^\mu - y_j^\mu) \sigma_i^\mu
	%	\begin{cases}
	%		w_{ji} - \eta \sigma_i^\mu, & \text{if } E_j < 0 \\
	%		
	%		w_{ji} + \eta \sigma_i^\mu, & \text{if } E_j > 0 \\ 
	%		
	%		0 ,&  \text{if } E = 0 ,
	%	\end{cases}
\end{equation}
%$E_i^\mu\equiv \xi_j^\mu - y_j^\mu $
where $\eta=0.001$ is the learning rate, $y_j^\mu$ is state of the output neuron $j$ without being clamped to the desired state $\xi_j^\mu$.  After all of the weights  $w_{ji}$ are updated, the pattern $\mu$ is removed, another pattern is presented, and the above procedure continues. The set of patterns are presented to the network for a number of times (epochs) and they are presented in random order in each epoch. After some number of presentations, it was checked whether the patterns are learned i.e. whether the patterns $\{\boldsymbol{\xi}^\mu\}$ are the fixed points of the network dynamics. A hard limit was imposed on the number of pattern presentations (5000 iterations). If after this maximum number of presentations, the patterns were not learned, the simulation was stopped, and learning the pattern set was considered unsuccessful. 
\\

The simulation of the mean-field approximation case is done as follows: for an arbitrary input unit $j$, we sample a Gaussian noise $z_{ij}^\mu$ for each pattern $\mu$ and each hidden unit $i$, and keep it fixed during learning. Each time the pattern $\mu$ is presented during learning, the hidden unit $i$ has the same value for the \textit{quenched} noise $z_{ij}^\mu$. In the main text we chose an arbitary $j$ then use notation $z_i^\mu \equiv z_{ij}^\mu$.
%We ran the simulation for 10 different realization of the network for a range of values of the pair $\alpha$ 

\section{Computing the probability distributions and correlations in the mean-field approximation model}
%\linenumbers
\subsection{General case}

As explained in the main text, after taking the MFA we can compute the probability distribution of the quantity ${\sigma_i^\mu= \text{sgn} \big(z_i^\mu+v_{ij}\xi_j^\mu\big)}$ i.e. the probability distribution of the hidden units. 
%The MFA model can now be reformulated as a perceptron problem for each $j$.

The conditional probability distributions of the hidden units given $\xi_j^\mu$ is	
\begin{align}
	\P\big(\sigma_i^\mu=1 \big\vert \xi_j^\mu=1\big)&=\P\big( z_i^\mu> -v_{ij}\big) \nonumber\\
	&=\frac{1}{\sqrt{2\pi N_v}}\int_{-v_{ij}}^{\infty} dx\, \me^{-\frac{x^2}{2N_v}} \nonumber\\
	% &=&\frac{1}{\sqrt{2\pi}} 
	&=\frac{1}{\sqrt{2\pi}}\int_{-\frac{v_{ij}}{\sqrt{N_v}}}^{\infty} d\hat {x}\,\me^{-\frac{\hat{x}^2}{2}}\nonumber\\
	&=H\big(-\frac{v_{ij}}{\sqrt{N_v}}\big)\nonumber\\
	&\simeq \frac{1}{2}+\frac{v_{ij}}{\sqrt{2\pi N_v}},\label{eq:prob1}
\end{align}
where change of variable $\hat{x}=\frac{x}{\sqrt{N_v}}$ is used and the notation $H(.)$ means the tail probability of the standard normal distribution. The last line is due to the asymptotic approximation of $H(.)$ to first order when $N_v\rightarrow\infty$. Similarly,

\begin{align}
	\P\big(\sigma_i^\mu=1 \big\vert \xi_j^\mu=-1\big)&=P\big( z_i^\mu> v_{ij}\big) 
	~~~~~~~~~~~~~\simeq~ \frac{1}{2}-\frac{v_{ij}}{\sqrt{2\pi N_v}}\label{eq:prob2}\\
	\P\big(\sigma_i^\mu=-1 \big\vert \xi_j^\mu=1\big)&=1-P\big(\sigma_i^\mu=1 \big\vert \xi_j^\mu=1\big) ~~\simeq~  \frac{1}{2}-\frac{v_{ij}}{\sqrt{2\pi N_v}}\label{eq:prob3}\\
	\P\big(\sigma_i^\mu=-1 \big\vert \xi_j^\mu=-1\big)&=1-P\big(\sigma_i^\mu=1 \big\vert \xi_j^\mu=-1\big) \simeq~  \frac{1}{2}+\frac{v_{ij}}{\sqrt{2\pi N_v}}\label{eq:prob4},
\end{align}
which can be written as
\begin{equation}
	\P\big(\sigma_i^\mu \big\vert \xi_j^\mu\big)\simeq\frac{1}{2}+\sigma_i^\mu\xi_j^\mu\frac{v_{ij}}{\sqrt{2\pi N_v}} \label{eq:prob4}.
\end{equation} 
It should be noted that the MFA makes the hidden neurons conditionally independent: 
\begin{equation}
	\P\big(\sigma_i^\mu \,\big\vert \, \xi_j^\mu\big) \perp \P\big(\sigma_k^\mu \, \big\vert \, \xi_j^\mu\big)\label{eq:perp}.
\end{equation}
Given this conditional distribution, it will be useful to compute the probability distribution of the quantity $\hat{\sigma}_i^\mu \equiv\sigma_i^\mu\xi_j^\mu$ for a fixed $j$ in the MFA as it will appear in the Gardner volume of solutions:
\begin{align}
	\P\big(\hat{\sigma}_i^\mu=1\big)&=\P\big(\sigma_i^\mu\xi_j^\mu=1\big)\nonumber\\
	%&=\P\big(\sigma_i^\mu=1,\xi_j^\mu=1\big)+\P\big(\sigma_i^\mu=-1,\xi_j^\mu=-1\big)\nonumber\\
	&=\P\big(\sigma_i^\mu=1 \,\big\vert \, \xi_j^\mu=1\big)\P\big(\xi_j^\mu=1\big)+\P\big(\sigma_i^\mu=-1 \,\big\vert \, \xi_j^\mu=-1\big)\P\big(\xi_j^\mu=-1\big)\nonumber\\
	&\simeq \Big(\frac{1}{2}+\frac{v_{ij}}{\sqrt{2\pi N_v}}\Big)\times\frac{1}{2}+\Big(\frac{1}{2}+\frac{v_{ij}}{\sqrt{2\pi N_v}}\Big)\times\frac{1}{2}\nonumber\\
	&=\frac{1}{2}+\frac{v_{ij}}{\sqrt{2\pi N_v}}\,,
	\label{sigmahat_1}
\end{align}
and similarly
\begin{align}
	\P\big(\hat{\sigma}_i^\mu=-1\big)=\P\big(\sigma_i^\mu\xi_j^\mu=-1\big)\simeq\frac{1}{2}-\frac{v_{ij}}{\sqrt{2\pi N_v}}\,.
	\label{sigmahat_-1}
\end{align}
Taking the definition 
\begin{equation}
	m_i\equiv \frac{2v_{ij}}{\sqrt{2\pi N_v}}
	\label{eq:def_m_i}
\end{equation}
we can now write the probability distribution of $\sigmahat_i^\mu$ as
\begin{equation}
	\P\big(\hat{\sigma}_i^\mu\big)=\frac{1}{2}(1+m_i)\,\delta\big(\hat{\sigma}_i^\mu-1\big)+\frac{1}{2}(1-m_i)\,\delta\big(\hat{\sigma}_i^\mu+1\big),
	\label{eq:prob_simgahat}
\end{equation}
where $\delta(.)$ is the Dirac delta function, so that $\llangle\hat{\sigma}_i^\mu\rrangle= m_i$ and the $\hat{\sigma}_i^\mu$'s are independent random variables:
\begin{equation}
	\P(\hat{\sigma}_i^\mu) \perp \P(\hat{\sigma}_k^\mu).
	\label{perp_sigmahat}
\end{equation}

\subsection{Sparse case}

We can also add sparseness in the hidden units representation (but patterns are dense), such that
\begin{equation}
	\P(\sigma_i^\mu = +1)= f\text{ and }\P(\sigma_i^\mu = -1)= 1-f.
	\label{sparseness}
\end{equation}
The output layer representation is not modified and kept at the dense regime, i.e. ${\P(\xi_j^\mu = \pm 1) = \frac{1}{2}}$.
The local field $\sum_{j=1}^{N_v} v_{ij} \xi_j^\mu$ at a hidden unit $\sigma_j^\mu$ is sampled from a Gaussian distribution of mean zero and deviation $\sqrt{N_v}$. We want to define a threshold $\theta$ such that
\begin{equation}
	{\int_{\theta}^{\infty}\dfrac{dt}{\sqrt{2 \pi N_v}}e^{\frac{-t^2}{2 N_v}}= f \int_{-\infty}^{\infty}\dfrac{dt}{\sqrt{2 \pi N_v}}e^{\frac{-t^2}{2 N_v}}} \text{, therefore }\theta \equiv \sqrt{N_v} H^{-1}(f).
\end{equation}
Taking the MFA, the hidden units become
\begin{equation}
	\sigma_{i\text{, sparse}}^\mu= \text{sgn}(z_i^\mu - \theta + v_{ij}\xi_j^\mu).
\end{equation}
The conditional probabilities of a hidden unit given $\xi_j^\mu$ can be computed again as
\begin{align}
	\P\big(\sigma_{i\text{, sparse}}^\mu &= +1 \big\vert \xi_j^\mu\big)\simeq f + \xi_j^\mu\frac{v_{ij}}{\sqrt{2\pi N_v}}\exp\left( - \frac{[H^{-1}(f)]^2}{2}\right) \label{eq:prob_sparse+} \\
	\P\big(\sigma_{i\text{, sparse}}^\mu &= -1 \big\vert \xi_j^\mu\big)\simeq 1 - f - \xi_j^\mu\frac{v_{ij}}{\sqrt{2\pi N_v}}\exp\left( - \frac{[H^{-1}(f)]^2}{2}\right)
	\label{eq:prob_sparse-}.
\end{align} 
Equations~\eqref{sigmahat_1} and \eqref{sigmahat_-1} become in the sparse case the probability distribution of ${\hat{\sigma}_{i\text{, sparse}}^\mu \equiv \xi_j^\mu \sigma_{i\text{, sparse}}^\mu}$
\begin{align}
	\P\big(\hat{\sigma}_{i\text{, sparse}}^\mu &= +1\big)\simeq\frac{1}{2}+ \frac{v_{ij}}{\sqrt{2\pi N_v}}\exp\left( - \frac{[H^{-1}(f)]^2}{2}\right) \\
	\P\big(\hat{\sigma}_{i\text{, sparse}}^\mu &= -1\big)\simeq\frac{1}{2}- \frac{v_{ij}}{\sqrt{2\pi N_v}}\exp\left( - \frac{[H^{-1}(f)]^2}{2}\right).
\end{align} 
This time we define
\begin{equation}
	\tilde{m}_i\equiv \frac{2v_{ij}}{\sqrt{2\pi N_v}}\exp\left( - \frac{[H^{-1}(f)]^2}{2}\right).
	\label{def_mtilde}
\end{equation}
The probability distribution in Eq.~\eqref{eq:prob_simgahat} becomes for $\hat{\sigma}_{i\text{, sparse}}^\mu$
\begin{equation}
	\P\big(\hat{\sigma}_{i\text{, sparse}}^\mu\big)=\frac{1}{2}(1+\tilde{m}_i)\,\delta\big(\hat{\sigma}_{i\text{, sparse}}^\mu-1\big)+\frac{1}{2}(1-\tilde{m}_i)\,\delta\big(\hat{\sigma}_{i\text{, sparse}}^\mu+1\big).
\end{equation}

\subsection{Input-output correlations in the MFA for the reduced perceptron problem}
The MFA reduces the problem to a capacity problem in a perceptron (for an arbitrary $j$) with input patterns and output labels $(\boldsymbol{\sigma}^\mu,\xi_j^\mu)$. The dependency between input and output is given by the conditional probability distribution Eq.~(\ref{eq:prob4}) in the general case. The simple pairwise correlation between input and output is ${\llangle \sigma_i^\mu \xi_j^\mu \rrangle = m_i}$.  We consider higher order input-output correlation of the form ${ \llangle \sigma_{i_1}^\mu \sigma_{i_2}^\mu...\sigma_{i_L}^\mu \xi_j^\mu \rrangle}$.
\begin{itemize}
	\item If $L$ is odd, using the independence of variables $\hat{\sigma}_i^\mu$'s stated in Eq.~\eqref{perp_sigmahat}:
\end{itemize}
\begin{equation}
	\llangle \left( \prod_{k=1}^{L} \sigma_{i_k}^\mu \right) \xi_j^\mu \rrangle = \llangle \prod_{k=1}^{L} \hat{\sigma}_{i_k}^\mu \rrangle = \prod_{k=1}^{L}  \llangle \hat{\sigma}_{i_k}^\mu \rrangle = \prod_{k=1}^{L} m_{i_k}.
\end{equation}
\begin{itemize}
	\item If $L$ is even, using the independence of the hidden units conditioned on $\xi_j^\mu$ stated\\ in Eq.~\eqref{eq:perp}:
\end{itemize}
\begin{align}
	\llangle \left( \prod_{k=1}^{L} \sigma_{i_k}^\mu \right) \xi_j^\mu \rrangle &= \llangle \left( \prod_{k=1}^{L} \sigma_{i_k}^\mu \right) \vert \xi_j^\mu = +1 \rrangle - \llangle \left( \prod_{k=1}^{L} \sigma_{i_k}^\mu \right) \vert \xi_j^\mu= -1 \rrangle \\
	&= \prod_{k=1}^{L} \llangle  \sigma_{i_k}^\mu  \vert \xi_j^\mu= +1 \rrangle - \prod_{k=1}^{L} \llangle  \sigma_{i_k}^\mu  \vert \xi_j^\mu= -1 \rrangle \\
	&= \prod_{k=1}^{L} m_{i_k} - \prod_{k=1}^{L} ( - m_{i_k}) \\
	&= \prod_{k=1}^{L} m_{i_k} - (-1)^L \prod_{k=1}^{L} m_{i_k} \\
	&= 0.
\end{align}
After taking the MFA, all higher-order input-output correlations are either null, or expressed as a product of pairwise correlations. In the full model, the correlation structure is more complex and richer, which explains why the capacity is higher in the full model that the capacity in the MFA.\\

In the case of sparseness in the hidden units and after taking the MFA, $m_i$ needs to be replaced by $\tilde{m}_i < m_i$, where $\tilde{m}_i$ decreases with sparseness. The input-output correlations have the same structure, but smaller values, which accounts for the decrease of capacity as the hidden units become more sparse (i.e. lowering coding level $f$). 

\section{Gardner analysis for the mean-field approximation (MFA) model}

%\subsection{The Gardner volume and the replica method}
We start out by enforcing the following spherical constraint for the weights $w_{ji}$ for fixed $j$

\begin{equation}
	\sum_{i=1}^{N}w_{ji}^2=N
	\label{eq:spherical}
\end{equation}
where we defined $N\equiv N_h$ for simplicity. As stated in the main text, we can enforce the spherical constraint on the encoding weights $\mathbf{v}_j\equiv v_{.j}$
\begin{equation}
	\sum_{i=1}^{N}v_{ij}^2=N,
	\label{eq:sphericalv}
\end{equation}
making the probability distribution of the encoding weights
\begin{equation}
	\P(\mathbf{v}_{j})=	(2\pi\me)^{-N/2}\delta(\|\mathbf{v}_{j}\|^2-N).
\end{equation}

In order for a pattern indexed $\mu$ to be a fixed point and be robust with a margin, we enforce the following requirement
\begin{equation}
	\xi_j^\mu\Big(\frac{1}{\sqrt{N}}\sum_{i=1}^{N}w_{ji}\sigma_i^\mu\Big) > \kappa  
	%\label{eq:mainconstraint}
	\label[cond]{cond:mainconstraint} %\Cref{cond:mainconstraint}
\end{equation}
where $\kappa$ is a robustness parameter providing a margin for the solution.

We are interested to compute, for our perceptron, the typical value of the Gardner volume that measures the subspace of solutions satisfying \Cref{cond:mainconstraint}  in the weight space for a given realization of input-output pair $\{\boldsymbol{\sigma}^\mu\}$ and $\{\xi_j^\mu\}$: 
\begin{equation}
	\Omega_\text{tot}(\boldsymbol{\sigma}^\mu,\xi_j^\mu)=\ddfrac{\int d^N\mathbf{w}\,\prod_{j=1}^{N_v} \delta\Big(\sum_{i=1}^{N}w_{ji}^2-N\Big) \prod_{j=1}^{N_v}\prod_{\mu=1}^{p} \Theta\Big( \xi_j^\mu\big(\frac{1}{\sqrt{N}}\sum_{i=1}^{N}w_{ji}\sigma_i^\mu\big) - \kappa \Big)}{\int d^N\mathbf{w}\,\prod_{j=1}^{N_v} \delta\Big(\sum_{i=1}^{N}w_{ji}^2-N\Big)}
	\label{eq:mainVol}
\end{equation}
where $\Theta(.)$ denotes the Heaviside step function. We observe that \Cref{eq:mainVol} factors into a product of identical terms for each $j$ so that $\Omega_\text{tot}=\prod_{j=1}^{N_v}\Omega_j$. 
Therefore, we study the following quantity
\begin{equation}
	\lim_{N\rightarrow\infty}\frac{1}{N}\log\Omega_\text{tot}=\frac{1}{N}\sum_j\log\Omega_j
\end{equation}
and we assume that it is self-averaging. So we only need to calculate $\llangle\log{\Omega}\rrangle$, the average of $\log\Omega_j$ over the quenched distributions of the patterns. To do that, we use the replica method
\begin{equation}
	\llangle\log{\Omega}\rrangle=\lim_{n\rightarrow 0}\frac{\llangle\Omega^n\rrangle-1}{n}
	\label{eq:replica}
\end{equation}
which assumes the validity of the analytical continuation from positive integer to real-values close to zero.

Now for simplicity we can drop the index $j$ all together and use the auxiliary variable $\hat{\sigma}_i^\mu \equiv\sigma_i^\mu\xi^\mu$, writing $\llangle\Omega^n\rrangle$ as
\begin{equation}
	\llangle\Omega^n\rrangle_{\boldsymbol{\sigmahat}^\mu,\mathbf{ v}}=\ddfrac{\Big\llangle\prod_{\alpha=1}^n\int d^N\mathbf{w}^\alpha\, \delta\Big(\sum_{i=1}^{N}(w_{i}^\alpha)^2-N\Big) \prod_{\mu=1}^{p} \Theta\Big( \frac{1}{\sqrt{N}}\sum_{i=1}^{N}w_{i}^\alpha\sigmahat_i^\mu - \kappa \Big)\Big\rrangle}  
	{\prod_{\alpha=1}^n\int d^N\mathbf{w}^\alpha\,\delta\Big(\sum_{i=1}^{N}(w_{i}^\alpha)^2-N\Big)},
	\label{eq:Volume_replica}
\end{equation} 
where the replicas are introduced with superscript notation $\alpha$.

%$\imath $

%\subsection*{Expansion of step functions}
%We use the following integral representation of the step function for their expansion
% \begin{equation}
% \Theta(z-\kappa)=\int_{\kappa}^{\infty}d\lambdah\delta(\lambdah-z)=\int_\kappa^\infty d\lambdah \int \frac{d\zhat}{2\pi}\me^{\icmplx \zhat(\lambdah-z)}
% \label{eq:integ_step}
% \end{equation}
% where $\icmplx=\sqrt{-1}$. 

Now let's denote the local field as %$z_\alpha^\mu$
\begin{align}
	z_\alpha^\mu %&= \xi^\mu N^{-1/2}\sum_i w_i^\alpha\sigma_i^\mu \\
	&= N^{-1/2}\sum_i w_i^\alpha\sigmahat_i^\mu
\end{align}
%  where $\sigmahat_i^\mu=\xi^\mu\sigma_i^\mu$ with probability distribution given by 
%    \begin{equation}
%    P\big(\hat{\sigma}_i^\mu\big)=\frac{1}{2}(1+m_i)\,\delta\big(\hat{\sigma}_i^\mu-1\big)+\frac{1}{2}(1-m_i)\,\delta\big(\hat{\sigma}_i^\mu+1\big),
%    	 \label{eq:prob_simgahat}
%    \end{equation}
%   and
and expand the step functions in \Cref{eq:Volume_replica} for each $\mu$ and $\alpha$ with their integral representation 
\begin{equation}
	%\Theta\Big(\xi^\mu N^{-1/2}\sum_i w_i^\alpha\sigma_i^\mu-\kappa\Big)=
	\Theta\Big(z_\alpha^\mu-\kappa\Big)=
	\int_{\kappa}^{\infty}\frac{d\lambdah_\alpha^\mu}{2\pi}
	\int d\zhat_\alpha^\mu\, \me^{\icmplx\zhat_\alpha^\mu\lambdah_\alpha^\mu}\
	\me^{-\icmplx\zhat_\alpha^\mu z_\alpha^\mu},
	\label{eq:step_localfield}
\end{equation}
where we introduced auxiliary variables $\lambdahat_\alpha^\mu$ and $\zhat_\alpha^\mu$.

Let's take the average of the last factor of \Cref{eq:step_localfield} over ${\boldsymbol{\sigmahat}^\mu}$
\begin{align}
	\Big\llangle\prod_{\mu\alpha}\me^{-\icmplx\zhat_\alpha^\mu z_\alpha^\mu}\Big\rrangle_{\boldsymbol{\sigmahat}^\mu} \label{averaging1}
	%&=\prod_{\mu\alpha}
	%\Big\llangle \exp\Big(-\icmplx\zhat_\alpha^\mu N^{-1/2}\sum_i w_i^\alpha\sigmahat_i^\mu\Big)\Big\rrangle\label{eq:step_localfield_last}\\
	&=\prod_{\mu i}
	\left\llangle \exp\Big(-\icmplx \sigmahat_i^\mu N^{-1/2}\sum_\alpha \zhat_\alpha^\mu w_i^\alpha\Big)\right\rrangle_{{\sigmahat}_i^\mu}\\
	%&=\Big\llangle \exp \Big(\sum_{\mu i}\log\Big[\exp\big(-\icmplx \sigmahat_i^\mu N^{-1/2}\sum_\alpha \zhat_\alpha^\mu w_i^\alpha\big)\Big]\Big)\Big\rrangle\\
	&=\exp \left\{\sum_{\mu i}\log\left[
	\ddfrac{1+m_i}{2}\exp\big(-\icmplx \sum_\alpha \ddfrac{w_i^\alpha}{\sqrt{N}}\zhat_\alpha^\mu \big)+
	\ddfrac{1-m_i}{2}\exp\big(\icmplx \sum_\alpha \ddfrac{w_i^\alpha}{\sqrt{N}}\zhat_\alpha^\mu \big) 
	\right]\right\}.
	\label{averaging2}
\end{align}
Expanding the exponentials and the logarithm to second order in $\sum_{\alpha}\ddfrac{w_i^\alpha\zhat_\alpha^\mu}{\sqrt{N}}$, considering the fact that other terms vanish in the thermodynamic limit, and using $ m_i = \frac{2v_{i}\sqrt{\Lambda}}{\sqrt{2\pi N}}$ give
\begin{equation}
	\Big\llangle\prod_{\mu\alpha}\me^{-\icmplx\zhat_\alpha^\mu z_\alpha^\mu}\Big\rrangle_{\boldsymbol{\sigmahat}^\mu}=\prod_{\mu}\exp\left(
	-\icmplx \frac{\sqrt{2\Lambda}}{\sqrt{\pi}}\sum_{\mu \alpha} M_\alpha \, \zhat_\alpha^\mu
	-\frac{1}{2} \Big( \sum_{\alpha} (\zhat_\alpha^\mu)^2
	+ 2 \sum_{\alpha<\beta} q_{\alpha \beta} \, \zhat_\alpha^\mu\, \zhat_\beta^\mu  \Big)
	\right),\label{eq:avgExp}
\end{equation}

where
\begin{align}
	M_\alpha &= \sum_i\ddfrac{v_i w_i^\alpha}{N}\label{eq:M},\\
	q_{\alpha \beta} &= \sum_i \ddfrac{w_i^\alpha w_i^\beta}{N},\label{eq:q_ab}
\end{align}
and the spherical constraint \Cref{eq:spherical} is used so that $q_{\alpha\alpha}=\sum_i (w_i^\alpha)^2/N=1$.

If we insert \Cref{eq:avgExp} into back into the integrals in the expansion of step functions, we see that we get an identical integral for each $\mu$, so we can drop the $\mu$'s obtaining
\begin{equation}
	\Big\llangle \Theta(z_\alpha^\mu -\kappa) \Big\rrangle_{\boldsymbol{\sigmahat}^\mu} 
	= \left[ \int_{\kappa}^{\infty} \Big( \prod_{\alpha} \ddfrac{d \lambdahat_\alpha}{2\pi}  \Big)
	\int \Big( \prod_{\alpha} d \zhat_{\alpha} \Big) \me^
	{K\{\lambdahat,\zhat, q, M\} }
	\right]^p,
	\label{eq:result_expand_Step}
\end{equation}
where 
\begin{equation}
	K\{\lambdahat,\zhat, q, M\} =  \icmplx \sum_\alpha \zhat_\alpha \lambdahat_\alpha %corrected mistake: -  to +
	- \icmplx \sqrt{2\Lambda/\pi} \sum_\alpha M_\alpha \zhat_\alpha
	- \frac{1}{2} \sum_\alpha \zhat_\alpha^2 
	- \sum_{\alpha<\beta} q_{\alpha\beta}\zhat_\alpha\zhat_\beta.
	\label{eq:K}
\end{equation}

%\subsection*{Expansion of $\delta$-functions}
%We use the following basic integral representation of the $\delta$-function
%  \begin{equation}
%	  \delta(z) = \int \ddfrac{dr}{2\pi \icmplx}\me^{-rz}.
%  \end{equation}
%Choosing $r=E_\alpha/2$ for each $\alpha$ in the delta functions in \Cref{eq:Volume_replica} yields
Similarly, we use the integral representation of the $\delta$-function to expand them in \Cref{eq:Volume_replica}
\begin{equation}
	\delta\Big( \sum_i (w_i^\alpha)^2 - N \Big) 
	= \int \ddfrac{d E_\alpha}{4\pi\icmplx} \, \me^{N E_\alpha/2 - E_\alpha \sum_i (w_i^\alpha)^2/2}.
\end{equation}
Also, we impose the condition in \Cref{eq:q_ab} for each pair of $\alpha,\beta$ (with $\alpha<\beta$) %by using $r=N F_{\alpha\beta}$ in the integral representation of the $\delta$-function:
\begin{equation}
	\delta\Big( q_{\alpha\beta} - \ddfrac{1}{N} \sum_i w_i^\alpha w_i^\beta  \Big)
	= N \int \ddfrac{dF_{\alpha\beta}}{2\pi\icmplx} \, \me^{ -N F_{\alpha\beta} q_{\alpha\beta} + F_{\alpha\beta}\sum_i w_i^\alpha w_i^\beta}.
\end{equation}
so that by integrating over each of the $q_{\alpha\beta}$'s, the delta function can pick out the desired value. Similarly, the constraint of \Cref{eq:M} can be imposed by
\begin{equation}
	\delta\Big( M_{\alpha} - \ddfrac{1}{N} \sum_i v_i w_i^\alpha  \Big)
	= N \int \ddfrac{d\Mhat_{\alpha}}{2\pi\icmplx} \, \me^{ -N \Mhat_{\alpha} M_{\alpha} + \Mhat_{\alpha}\sum_i v_i w_i^\alpha}.
\end{equation}

We can now factorize the integrals over $\mathbf{w}$ in \Cref{eq:Volume_replica}. Considering factors involving $w_i^\alpha$, the numerator of \Cref{eq:Volume_replica} includes the following integral
\begin{align}
	\int \Big(\prod_{\alpha i} d w_i^\alpha \Big) \me^{-\sum_{\alpha,i} E_\alpha (w_i^\alpha)^2/2 
		+ \sum_{\alpha<\beta,i}F_{\alpha\beta} w_i^\alpha w_i^\beta
		+
		\sum_{\alpha,i} \Mhat_{\alpha} v_i w_i^\alpha
	}.
\end{align}

This is the only place that the index $i$ appears in the numerator. Thus, we drop the index $i$ in $w_i$ and rewrite the above integral as multiplication of $N$ integrals
\begin{equation}
	\prod_{i=1}^{N}
	%\left[ 
	\int \Big(\prod_{\alpha} d w^\alpha \Big) \me^{-\sum_\alpha E_\alpha (w^\alpha)^2/2 
		+ \sum_{\alpha<\beta}F_{\alpha\beta} w^\alpha w^\beta
		+
		\sum_\alpha \Mhat_{\alpha} v_i w^\alpha
	}
	%\right]
	.
\end{equation}
Following a similar calculation in the denominator, one obtains
\begin{equation}
	\left[ 
	\int \Big(\prod_{\alpha} d w^\alpha \Big) \me^{-\sum_\alpha E_\alpha (w^\alpha)^2/2 }
	\right]^N.
\end{equation}

%\subsection*{Replica-symmetric ansatz}
Now, we can collect all the terms together writing the volume in \Cref{eq:Volume_replica} as
\begin{equation}
	\llangle\Omega^n\rrangle_{\boldsymbol{\sigmahat}^\mu,\mathbf{v}} 
	= \left\llangle \ddfrac{\int \Big( \prod_{\alpha} d E_\alpha \Big)
		\Big( \prod_{\alpha} d \Mhat_\alpha \Big) 
		\Big( \prod_{\alpha<\beta} dq_{\alpha\beta} \, dF_{\alpha\beta} \Big)
		\me^{ N \, G\{q_{\alpha \beta},M_\alpha,E_\alpha,F_{\alpha \beta}\}} }
	{\int \Big( \prod_{\alpha} d E_\alpha \Big) \me^{N \tilde{G}\{ E_\alpha \}} }\right\rrangle_\mathbf{v},
	\label{eq:volume2}
\end{equation}
where the `free energy' $G$ and the $\tilde{G}$ are
\begin{flalign}
	G\{q_{\alpha \beta},M_\alpha,\Mhat_\alpha,E_\alpha,F_{\alpha \beta}\} &=\alpha G_1\{q_{\alpha \beta},M_\alpha\} + G_2\{E_\alpha,F_{\alpha \beta},\Mhat_\alpha\} 
	- \sum_{\alpha<\beta} F_{\alpha\beta} \, q_{\alpha\beta} + \frac{1}{2} \sum_\alpha E_\alpha -
	\sum_\alpha \Mhat_\alpha M_\alpha
	&&
	\label{eq:G}\\
	G_1\{q_{\alpha \beta},M_\alpha\}   &= \log\left[\int_{\kappa}^{\infty}
	\Big( \prod_\alpha \ddfrac{d\lambdahat_\alpha}{2\pi} \Big)
	\int \Big( \prod_\alpha d \zhat_\alpha \Big)
	\,\me^{%\Big( 
		\icmplx \sum_\alpha \zhat_\alpha \lambdahat_\alpha %corrected: - to +
		- \icmplx \sqrt{2\Lambda/\pi} \sum_\alpha M_\alpha \zhat_\alpha
		- \frac{1}{2} \sum \zhat_\alpha^2 
		- \sum_{\alpha<\beta} q_{\alpha\beta}\zhat_\alpha\zhat_\beta
		%\Big)
	}
	\right]&&\\
	G_2\{E_\alpha,F_{\alpha \beta},\Mhat_\alpha\} &=\frac{1}{N}\log 
	\left[
	\prod_{i=1}^{N} \int \Big( \prod_{\alpha} d w_\alpha \Big)
	\, \me^{-\frac{1}{2}\sum_\alpha E_\alpha w_\alpha^2 + \sum_{\alpha<\beta} F_{\alpha\beta}w_\alpha w_\beta +
		\sum_\alpha \Mhat_{\alpha} v_i w_\alpha
	}	  
	\right] &&\\
	\tilde{G}\{ E_\alpha \}  &= \log \left[
	\int \Big( \prod_{\alpha} d w_\alpha \Big)
	\, \me^{-\frac{1}{2}\sum_\alpha E_\alpha w_\alpha^2
	}			  
	\right] + \frac{1}{2} \sum_{\alpha}E_\alpha,
\end{flalign}
where $\alpha=p/N$ is the capacity variable.

We note that the exponents inside the integrals in \Cref{eq:volume2} are proportional to $N$, therefore we will be able to evaluate them in the large-$N$ limit using the saddle-point method over $F_{\alpha\beta}$, $q_\alpha$, $M_\alpha$, $\Mhat_\alpha$, and $E_\alpha$. In order to find this saddle point, we make the replica-symmetric (RS) ansatz
\begin{equation} %or use \nonumber in align environment
	\label{eq:RS}
	\begin{aligned}
		q_{\alpha\beta} &= q && \alpha < \beta \\
		F_{\alpha\beta} &= F && \alpha < \beta \\
		M_\alpha &= M        && \text{for all } \alpha \\	  
		\Mhat_\alpha &= \Mhat        && \text{for all } \alpha \\
		E_\alpha &= E        && \text{for all } \alpha.
	\end{aligned}
\end{equation}		  
Since the space of solutions is connected, the RS assumption is reasonable. We will show at the end of this supplementary text that the RS solutions are indeed locally stable.

The RS ansatz allows us to calculate each term in $G$. 
The integral over $\zhat_\alpha$ in $G_1$ can be done using the Gaussian integral trick and  the replica trick in the limit of $n\rightarrow0$ yielding
\begin{align}
	G_1 
	%&= \log\left[\int_{\kappa}^{\infty}
	%		 \Big( \prod_\alpha \frac{d\lambdahat_\alpha}{2\pi} \Big)
	%		 \int \Big( \prod_\alpha d \zhat_\alpha \Big)
	%		 \int \Dx{t}{}
	%		 \,\me^{
	%			 \icmplx (t\sqrt{q} +\lambdahat_\alpha - \sqrt{2\Lambda/\pi}\,M ) \sum_\alpha \zhat_\alpha %corrected: -\lambdahat_\alpha to +\lambdahat_\alpha
	%			\label{key}- \frac{1-q}{2}  \sum_\alpha \zhat_\alpha^2
	%		 }
	%		 \right]\\
	%		 & =  \log\left[\int \Dx{t}{}\left(
	%		 \int_{\kappa}^{\infty}
	%		  \frac{d\lambdahat}{2\pi}
	%		  \int d \zhat		 
	%		 \,\me^{
	%		 	 \icmplx (t\sqrt{q} + \lambdahat - \sqrt{2\Lambda/\pi}\,M ) \zhat %corrected: -\lambdahat to +\lambdahat
	%		 	\label{key}- \frac{1-q}{2}  \zhat^2
	%		 }
	%		 \right)^n
	%		 \right]\\
	%		 & =  \log\left[\int \Dx{t}{}\left(	
	%		 \int_{\kappa}^{\infty} \frac{d \lambdahat}{\sqrt{2\pi(1-q)}} \me^{-\frac{\left(t\sqrt{q}+ \lambdahat - \sqrt{2\Lambda/\pi}\,M \right)^2}{2(1-q)}} %corrected: -\lambdahat to +\lambdahat
	%		 \right)^n
	%		 \right]\\
	= n\int
	\Dx{t}{}
	\log
	\int_{\kappa}^{\infty} \frac{d \lambdahat}{\sqrt{2\pi(1-q)}} \me^{-\frac{\left(t\sqrt{q}+ \lambdahat - \sqrt{2\Lambda/\pi}\,M \right)^2}{2(1-q)}}, 
\end{align}
where $Dt \equiv \ddfrac{dt}{\sqrt{2\pi}}\me^\frac{-t^2}{2}$.
Similarly, $G_2$ can be calculated in the limit of $n\rightarrow 0$ yielding
\begin{align}
	G_2 
	=\frac{n}{2}
	\left(
	\log 2\pi - \log(E+F) + \frac{F + \Mhat^2 (\frac{1}{N}\sum_{i=1}^{N} v_i^2)}{E+F}		
	\right).
\end{align}
Averaging out $v$ which makes $\frac{1}{N}\sum_{i=1}^N v_i^2 =1$
% (considering \Cref{eq:sphericalv}) 
yields
\begin{align}
	G_2 = \frac{n}{2} \left(
	\log 2\pi - \log(E+F) + \frac{F + \Mhat^2}{E+F}		
	\right)
\end{align}
Now, inserting $G_1$ and $G_2$ into \Cref{eq:G} yields
\begin{align}
	G(q,M,E,F) &= \alpha \, n\int
	\Dx{t}{}
	\log \left[
	\int_{\kappa}^{\infty} \frac{d \lambdahat}{\sqrt{2\pi(1-q)}} \me^{-\frac{\left(t\sqrt{q}+ \lambdahat - 	\sqrt{2\Lambda/\pi}\,M \right)^2}{2(1-q)}} %corrected: -\lambdahat to +\lambdahat
	\right]\\
	&+ \frac{n}{2}\big(\log 2\pi - \log(E+F) + \frac{F+ \Mhat^2}{E+F}\big)\\
	&+ \frac{n}{2}(E+qF-2\Mhat M).
\end{align}
%We now turn to finding the saddle point of G with respect to $q$, $M$, $\Mhat$, $E$, and $F$, then taking the limit of $q\rightarrow 1$. %TODO: explain why

%\subsection*{Saddle-point equations}
Solving the saddle point equations $\partial G/\partial \Mhat = 0$, $\partial G/\partial F = 0$ and $\partial G/\partial E = 0$ yield
\begin{align}
	\Mhat &= \frac{M}{1-q} \\
	F     &= \frac{q-M^2}{(1-q)^2}\\
	E     &= \frac{1-2q+M^2}{(1-q)^2}.	  
\end{align}
Inserting these into $G$ gives %and using  change of variable $\tau=\frac{\lambdahat+t\sqrt{q}-M\sqrt{2\Lambda/\pi}}{\sqrt{1-q}}$ give
\begin{align}
	G(q,M) &= n \,\alpha \int \Dx{t}{} \log 
	H\big(\frac{\kappa + t\sqrt{q} - M\sqrt{2\Lambda/\pi}}{\sqrt{1-q}}\big)
	%\left[
	%\int_{\frac{\kappa + t\sqrt{q} - M\sqrt{2\Lambda/\pi}}{\sqrt{1-q}}}^{\infty} \Dx{\tau}{} %fixed
	%\right]
	+ \frac{n}{2} \big(
	\log 2\pi + \log(1-q) + \frac{q}{1-q} + 1 - \frac{M^2}{1-q}
	\big),&&
\end{align}
where $H(x)\equiv \frac{1}{\sqrt{2\pi}}\int_{x}^{\infty} d\tau\,\me^{-\frac{\tau^2}{2}}$.
Setting $\partial G/\partial q = 0$ yields
\begin{equation}
	\alpha \int \Dx{t}{}
	\,[H(u)]^{-1}
	%\left[
	%\int_{u}^{\infty} \Dx{\tau}{}
	%\right]^{-1}
	\frac{1}{\sqrt{2\pi}}\me^{-\frac{u^2}{2}}\,
	\Big(\frac{t+\kappa\sqrt{q}-M\sqrt{2\Lambda/\pi}\sqrt{q}}{2\sqrt{q}(1-q)^{3/2}}\Big)
	= \frac{q-M^2}{2(1-q)^2},
	\label{eq:dGdq}
\end{equation}
where $u=\frac{\kappa + t\sqrt{q} - M\sqrt{2\Lambda/\pi}}{\sqrt{1-q}}$.
The equation $\partial G/\partial M=0$ gives
\begin{align}
	\alpha \int \Dx{t}{}
	\,[H(u)]^{-1}
	%\left[
	%\int_{u}^{\infty} \Dx{\tau}{}
	%\right]^{-1}
	\frac{1}{\sqrt{2\pi}}\me^{-\frac{u^2}{2}}\,
	\Big(\frac{\sqrt{2\Lambda/\pi}}{\sqrt{1-q}}\Big)
	= \frac{M}{1-q}.
	\label{eq:dGdM}
\end{align}
Taking the limit of $q \rightarrow 1$ in \Cref{eq:dGdq} and (\ref{eq:dGdM})  yields the following two integral equations for critical capacity $\alpha_c$ and $M$ as a function of $\Lambda$ for the MFA model:
\begin{align}
	\alpha_c \int_{M\sqrt{2 \Lambda/\pi} - \kappa}^{\infty} Dt \, \big( \kappa + t - \sqrt{2 \Lambda/\pi} \big)^{2} = 1- M^{2}\label{eq:result1}\\
	\alpha_c \int_{M\sqrt{2 \Lambda/\pi}-\kappa}^{\infty} Dt\, \big( \kappa + t - \sqrt{2 \Lambda/\pi} \big) \sqrt{2 \Lambda/\pi}=M,\label{eq:result2}
\end{align}
which can be solved numerically and is plotted in the paper.
\\

In the sparse case, averaging in Eq.~\eqref{averaging1} yields the same result as Eq.~\eqref{averaging2}, but after replacing $m_i$ by $\tilde{m}_i$ defined in Eq.~\eqref{def_mtilde}:
\begin{align}
	\Big\llangle\prod_{\mu\alpha}\me^{-\icmplx\zhat_\alpha^\mu z_\alpha^\mu}\Big\rrangle_{\boldsymbol{\sigmahat}_{\text{sparse}}^\mu} 
	&=\prod_{\mu i}
	\left\llangle \exp\Big(-\icmplx \sigmahat_{i\text{, sparse}}^\mu N^{-1/2}\sum_\alpha \zhat_\alpha^\mu w_i^\alpha\Big)\right\rrangle_{{\sigmahat}_{i\text{, sparse}}^\mu}\\
	&=\exp \left\{\sum_{\mu i}\log\left[
	\ddfrac{1+\tilde{m}_i}{2}\exp\big(-\icmplx \sum_\alpha \ddfrac{w_i^\alpha}{\sqrt{N}}\zhat_\alpha^\mu \big)+
	\ddfrac{1-\tilde{m}_i}{2}\exp\big(\icmplx \sum_\alpha \ddfrac{w_i^\alpha}{\sqrt{N}}\zhat_\alpha^\mu \big) 
	\right]\right\}.
\end{align}
This leads us to defining
\begin{equation}
	\tilde{M}_\alpha = \sum_i\ddfrac{v_i w_i^\alpha}{N} \exp\left( - \frac{[H^{-1}(f)]^2}{2}\right)= \exp\left( - \frac{[H^{-1}(f)]^2}{2}\right)M_\alpha,
\end{equation}
which replaces $M_\alpha$ in the averaged volume in Eq.~\eqref{eq:result_expand_Step}. All previously derived equations remain the same in the sparse case, after replacing $M_\alpha$ by $\tilde{M}_\alpha$. The result in the MFA with sparseness in the hidden units representations finally read
\begin{align}
	\alpha_c \int_{\tilde{M}\sqrt{2 \Lambda/\pi} - \kappa}^{\infty} Dt \, \big( \kappa + t - \sqrt{2 \Lambda/\pi} \big)^{2} = 1- \tilde{M}^{2}\label{eq:result1}\\
	\alpha_c \int_{\tilde{M}\sqrt{2 \Lambda/\pi}-\kappa}^{\infty} Dt\, \big( \kappa + t - \sqrt{2 \Lambda/\pi} \big) \sqrt{2 \Lambda/\pi}=\tilde{M}\label{eq:result2}.
\end{align}

\section{Stability of the replica-symmetric solution}

We also performed stability analysis showing the replica-symmetric solution is locally stable, following Gardner's analysis \cite{gardner88}. The local stability is determined by the eigenvalues of the matrix $\mathcal{M}$ of quadratic fluctuations of $G\{q_{\alpha \beta},M_\alpha,\Mhat_\alpha,E_\alpha,F_{\alpha \beta}\}$ given in Eq.~\eqref{eq:G} in the variables $q_{\alpha \beta}$, $M_\alpha$, $\Mhat_\alpha$, $E_\alpha$, $F_{\alpha \beta}$ at the saddle-point. In an appropriate basis, we write: \[\mathcal{M}= \left( \begin{array}{ccccc}
A & -I_{n(n-1)/2} & & & \\
-I_{n(n-1)/2} & B & & & \\
&  &  C & -I_{n} & \\
&  &  -I_{n} & D & \\
&  &  &  &  E \end{array}  \right)\]
${A= \left( \dfrac{\partial^2 G_1}{\partial q_{\alpha \beta} \partial q_{\gamma \delta}}\right)_{\alpha < \beta, \gamma < \delta}}$, ${B= \left( \dfrac{\partial^2 G_2}{\partial F_{\alpha \beta} \partial F_{\gamma \delta}}\right)_{\alpha < \beta, \gamma < \delta}}$, ${C= \left( \dfrac{\partial^2 G_1}{\partial M_{\alpha} \partial M_{\beta}}\right)_{\alpha, \beta}}$, ${D= \left( \dfrac{\partial^2 G_2}{\partial \hat{M}_{\alpha} \partial \hat{M}_{\beta}}\right)_{\alpha, \beta}}$, ${E= \left( \dfrac{\partial^2 G_2}{\partial E_{\alpha} \partial E_{\beta}}\right)_{\alpha, \beta}}$.
$\mathcal{M}$ can be split into two blocks $\mathcal{M}_{1}= \left( \begin{array}{cc}
A & -I_{n(n-1)/2} \\
-I_{n(n-1)/2} & B\end{array} \right)$ and $\mathcal{M}_{2}= \left( \begin{array}{ccc}
C & -I_{n} & \\
-I_{n} & D & \\
& & E\end{array} \right)$.\\ $\mathcal{M}$'s spectrum is made by $\mathcal{M}_1$ and $\mathcal{M}_2$'s eigenvalues. We first focus on $\mathcal{M}_1$. To determine $\mathcal{M}_1$'s eigenvalues, we will rely on the replica-symmetric properties of the matrix and follow three steps: computing the elements of $A$ and $B$, finding their eigenvalues, and finally deducing the eigenvalues of $\mathcal{M}_1$. We reproduce the detailed explanation of Appendix 4 in \cite{engel01}.
\begin{itemize}
	\item Elements of $A$ and $B$
\end{itemize}
Given the structure of $A$ and $B$, each one of those matrices has only three elements. For instance 
$A_{\alpha \beta, \gamma \delta}= \left\{
\begin{array}{lll}
P & \mbox{if } \alpha = \gamma, \beta = \delta \\
Q & \mbox{if two indices coincide} \\
R & \mbox{if all indices are different from each other.}
\end{array}
\right.$
\begin{itemize}
	\item Eigenvalues of $A$ and $B$
\end{itemize}
The eigenvalues of $A$ and $B$ are almost entirely determined by the replica-symmetric properties of these matrices. Taking $A$, we can show that it has three eigenvalues, and only one of them can change sign and reflect local instability of the replica symmetry. The significant eigenvalue can be expressed as a function of $A$'s elements, and at the saddle-point in the limit $q \rightarrow 1$ it becomes
\begin{equation}
	\lambda_{A}=P - 2Q + R= \dfrac{\alpha_c}{(1-q)^{2}} \int_{M\sqrt{2\Lambda/ \pi} - \kappa}^{\infty} Dt
	\label{lambdaA}.
\end{equation}
Similarly, the only significant eigenvalue of $B$ when $q\rightarrow 1$ is
\begin{equation}
	\lambda_{B}=(1-q)^2
	\label{lambdaB}.
\end{equation}
\begin{itemize}
	\item Eigenvalues of $\mathcal{M}_1$
\end{itemize}
One can show that $\mathcal{M}_1$'s eigenvalues satisfy the equation ${X^2 - (\lambda_A + \lambda_B) X + \lambda_A \lambda_B - 1 = 0}$. To determine the local stability of the saddle-point, we consider $\mathcal{M}_1$'s two eigenvalues $X_1$ and $X_2$, which should have the same sign. Their product is given by
\begin{equation}
	X_1 X_2 = \lambda_A \lambda_B - 1.
	\label{product-ev}
\end{equation}
When $\alpha=0$ the product in Eq.~\eqref{product-ev} of the eigenvalues is $-1$, and the solution is stable in this limit as the Gardner volume is simply an integral over the phase space of weights. This apparently "wrong" negative sign of $X_1 X_2$ is due to the change of variable $F \rightarrow iF$ when introducing $F$ as the conjugate variable of $q$. To guarantee stability, the sign of the eigenvalues should not change, hence the quantity in Eq.~\eqref{product-ev} should remain negative. In the limit $q \rightarrow 1$, using the saddle-point Eq.~ \eqref{eq:result1} and \eqref{eq:result2}, we find
\begin{equation}
	X_1 X_2 = \alpha_c \int_{M\sqrt{2\Lambda/ \pi} - \kappa}^{\infty} Dt - 1 = - \kappa M \sqrt{\dfrac{2 \pi}{\Lambda}} < 0.
\end{equation}
We have therefore shown stability of $\mathcal{M}_1$ at critical capacity. We can also compute $\mathcal{M}_2$'s eigenvalues, which are of order 1 and go to zero when $q \rightarrow 1$, and are thus negligible with respect to $\mathcal{M}_1$'s eigenvalues. Stability is entirely determined by $\mathcal{M}_1$, and we find that the replica-symmetric solution is locally stable.

\section{Towards robustness to noise in the input layer}

In this part, we make a preliminary attempt to characterize the robustness to one bit flip in the input layer. We flip an arbitrary input unit $j$, and adapt the result from Eq.~(12) in \cite{babadi14} in the limit ${ \Delta S  = \frac{2}{N_v} \ll 1}$ and ${f=\frac{1}{2}}$ finding the average absolute change in the hidden units is $\langle \mid \Delta \sigma^\mu \mid \rangle= \frac{2}{\pi \sqrt{N_v}}$. We know the change in one hidden unit $\sigma_i^\mu$ is discrete, and takes its value from $\{ -2, 0, 2\}$. Besides, after taking the MFA, the conditional probabilities of the $\sigma_i^\mu$'s given $\xi_j^\mu$ are independent. It can be shown that the contribution of the $ \mid \Delta \sigma_i^\mu \mid $'s to $ \mid \Delta \sigma^\mu \mid  $ are independent, and each $\mid \Delta \sigma_i^\mu \mid$ is linear with $v_{ij}$. We can deduce that the change of one hidden unit $\sigma_i^\mu$ has the following probability distribution:
%Knowing $\Delta \sigma^\mu$ and the linearity of the local field in the $v_{ij}$'s, we can deduce that the change of one hidden unit $\sigma_i^\mu$ has the following probability distribution:
\begin{equation}
	\P (\Delta \sigma_i^\mu) = \dfrac{\mid v_{ij} \mid}{\pi \sqrt{N_v}}  \delta (\Delta \sigma_i^\mu - 2) + \dfrac{ \mid v_{ij} \mid}{\pi \sqrt{N_v}} \delta (\Delta \sigma_i^\mu +2 ) + \big( 1 - \dfrac{2 \mid v_{ij} \mid }{\pi \sqrt{N_v}} \big) \delta (\Delta \sigma_i^\mu).
\end{equation}
We now consider the local field $h_j$ at the output unit $j$, and we are interested in its change $\Delta h_j = \frac{1}{\sqrt{N_h}} \sum_{i=1}^{N_h} w_{ji} \Delta \sigma_i^\mu$ due to one bit flip in the input layer. We evaluate this change in the case where $v_{ij} > 0$ for all $i$, and $\xi_j^\mu$ is flipped from $-1$ to $+1$, i.e. $\Delta \sigma_i^\mu$ can only take its value from $\{ 0, 2\}$. The average change in the local field at unit $j$ is then
\begin{equation}
	\langle \Delta h_j \rangle_{(\Delta \sigma_i^\mu \vert v_{ij} >0,\text{ } \xi_j^\mu: -1 \rightarrow +1)} = \dfrac{1}{\sqrt{N_h}} \sum_{i=1}^{N_h} w_{ji}  \dfrac{2 v_{ij}}{\pi \sqrt{N_v}} = \dfrac{2 \sqrt{\Lambda} M}{\pi}.
\end{equation}
To be robust to one bit flip in the input layer, we should set the robustness parameter to $\kappa = \frac{2 \sqrt{\Lambda} M}{\pi}$. Plugging this value of $\kappa$ in Eqs.~\eqref{eq:result1} and \eqref{eq:result2} and solving them numerically, we find that the capacity now decreases slightly with the expansion ratio. Robustness to noise in the input layer needs further investigation. It remains to be seen whether the full-network has the same properties as the MFA model, and if learning the encoding weights or sparse connectivity could improve the trade-off between this robustness and the capacity of the network.

\bibliographystyle{abbrv}
%\bibliography{expcapacity}

\end{document}